%% file: ms.tex
\documentclass[11pt,a4paper]{article}
\usepackage[a4paper, total={18cm, 25cm}]{geometry}

\usepackage{currfile}
\usepackage[round]{natbib}
\usepackage{sources/latex_packages/myLatexStyle/myLatexStyleArticleNeutral/myStyle}
\usepackage{sources/paper-related-notations} 

\newcommand{\pathLatexSource}{sources}

\newcommand{\pathImagesOptimalTrading}{images}

\newcommand{\pathImagesTexOptimalTrading}{images/tikz/}

\newcommand{\pathTxtFilesOptimalTrading}{txt/}

\newcommand{\orcid}[1]{\href{https://orcid.org/#1}{\includegraphics[width=.4cm]{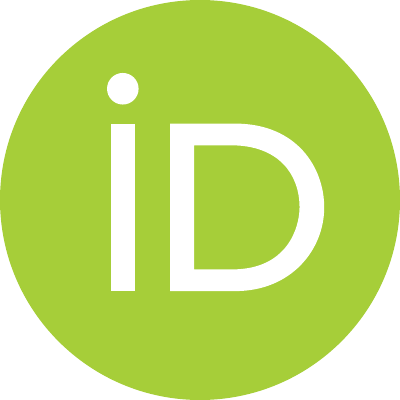}}}

\date{\monthyeardate\today}

\title{
  Optimal Trade Execution with Uncertain Volume Target
  \thanks{
    The work of the first author was supported by The Alan Turing Institute under the Turing Doctoral Studentship grant TU/C/000022.
    The work of the second author was supported by The Alan Turing Institute under the EPSRC grant EP/N510129/1.
  }
}

\author{
  Julien Vaes
  \thanks{
    Mathematical Institute,
    University of Oxford,
    Andrew Wiles Building,
    Radcliffe Observatory Quarter,
    Woodstock Road,
    Oxford,
    OX2 6GG 
    (\href{mailto:julien.vaes@maths.ox.ac.uk}{julien.vaes@maths.ox.ac.uk}).
  }
  \hspace{2mm}\orcid{0000-0003-4562-5373}%
  \and Raphael Hauser
  \thanks{
    Mathematical Institute,
    University of Oxford,
    Andrew Wiles Building,
    Radcliffe Observatory Quarter,
    Woodstock Road,
    Oxford,
    OX2 6GG 
    (\href{mailto:hauser@maths.ox.ac.uk}{hauser@maths.ox.ac.uk}).
  }
  \hspace{2mm}\orcid{0000-0002-1166-5329}%
}

\begin{document}

\maketitle

\begin{abstract}
\input{\pathSectionsOptimalTrading/0-abstract}
\end{abstract}


\input{\pathSectionsOptimalTrading/0-main.tex}


\bibliography{references}


\end{document}

%% file: sections/0-abstract.tex
\noindent
In the seminal paper on optimal execution of portfolio transactions,
\cite{Almgren2001} define the optimal trading strategy to liquidate a fixed volume of a single security under price uncertainty.
Yet there exist situations,
such as in the power market,
in which the volume to be traded can only be estimated and becomes more accurate when approaching a specified delivery time.
During the course of execution,
a trader should then constantly adapt their trading strategy to meet their fluctuating volume target.
In this paper,
we develop a model that accounts for volume uncertainty and we show that a risk-averse trader has benefit in delaying their trades.
More precisely,
we argue that the optimal strategy is a trade-off between early and late trades in order to balance risk associated with both price and volume.
By incorporating a risk term related to the volume to trade,
the static optimal strategies suggested by our model avoid the explosion in the algorithmic complexity usually associated with dynamic programming solutions,
all the while yielding competitive performance.

%% file: sections/0-main.tex
\input{\pathSectionsOptimalTrading/1-introduction}
\input{\pathSectionsOptimalTrading/2-models}

\input{\pathSectionsOptimalTrading/3-numerical_results}

\input{\pathSectionsOptimalTrading/4-comments}
\input{\pathSectionsOptimalTrading/6-conclusion}

%% file: sections/1-introduction.tex
\section{Introduction}
\label{sec:introduction}

The optimal execution problem refers to the problem of finding the best trading strategy in order to ensure the transition from one portfolio to another within an allocated period of time.
A trading strategy consists of buy and sell orders exchanged on the dedicated market.
This problem of optimal execution is germane to markets where liquidity is insufficient.
Indeed,
in insufficiently liquid markets,
price dynamics are sensitive to large trades.
Hence,
a rational trader should take into consideration the impact of their own trades,
which is assumed adverse to their trading position in the sense that it increases their trading cost.
In their seminal paper on optimal execution,
\citet{Almgren2001} proposed a model that captures both the inherent random evolution of the prices as well as the effect of trades on the price dynamics.
By adopting the return-risk trade-off defined by \citet{Markowitz1952} for portfolio hedging,
\citet{Almgren2001} added the missing nuance of risk to the optimal order execution problem previously stated in \cite{Bertsimas1998}.
They showed that,
the more risk-averse a trader, 
the quicker the acquisition of the desired position should be carried out in order to avoid the risk associated with the price dynamics.
These strategies are static \comadd{and time consistent:
They are determined in advance of trading in the sense that they only depend on the information available prior to the start of the execution period.
Also,
if a trader recomputes the optimal strategy mid-course,
the future trading plan remains unchanged since future price fluctuations are assumed independent of the past realisations.}
Nevertheless,  
if a trader defines as part of their initial strategy a mid-course updating rule,
\emph{aggressive-in-the-money} strategies in the sense of \citet{Kissell2005} lead to better return-risk trade-offs \citep{Almgren2007,Lorenz2011}.
As the variance penalises both the advantageous and adverse trading cost outcomes without any distinction,
other risk measures have also been used in the literature,
such as the \textit{expected utility} 
\citep{Schied2009,Schoneborn2016},
or the \textit{$\alpha$-Conditional Value-at-Risk} ($\CVaR_\alpha$) 
\citep{Butenko2005,Feng2012,Krokhmal2007}.

In the papers previously mentioned,
the total volume of securities to be traded is deterministic and given as a parameter.
However,
in many situations,
a trader only disposes of an estimate of this volume,
which varies throughout the course of execution;
there is thus need for a model that combines the risk associated with both the price dynamics and the uncertainty of the volume target.
\comrefo{
For example, the manager of an open ended investment fund may issue or redeem shares to avoid the asset-liability mismatch that could ensue 
if shares were only traded in the secondary market. Orders are usually collected and aggregated during the trading day, and the issue or 
redemption of the net share order creates the need to acquire or dispose of a certain volume of the fund's investment assets, typically over 
the next two trading days. Our model allows the fund manager to front load some of these trades and execute orders closer to the time when 
they are first placed, before all share orders are netted out at the end of the day, thus greatly increasing the market liquidity for the 
fund's investors. 
}
Another example is the clearing of power futures markets,
which has to result in an outstanding volume close to the realised demand in every given delivery period,
as the spot market has limited liquidity due to the physical constraints of the power plants used for generation.

To our knowledge,
no previous model considers both of these two sources of uncertainty.
Related articles are \citet{Cheng2017,Bulthuis2017,Cheng2019},
where the optimal strategy with uncertain order fills is investigated,
\ie the risk for an order to be filled either incompletely or in excess (the latter being considered mainly for mathematical than practical reasons). 
In significant contrast to our situation,
in their setting,
the magnitude of the uncertainty of order fills is assumed to be either constant or proportional to the order size.
In our case,
the uncertainty is independent of the trader's decisions:
the updates in the volume forecast depend exclusively on extraneous variables.
Our model assumes consequently that both sources of uncertainty, 
\ie the price dynamics and the forecast updates of the volume target,
are inherent to the market and independent of the trader's trading strategy.
\comadd{	
Moreover,
the model presented in this paper allows for more flexibility compared to the approaches adopted by
\citet{Cheng2017,Bulthuis2017,Cheng2019}
as it does not rely on the assumption that the trading periods are homogeneous:
for instance,
it allows for the asset's liquidity to vary over the course of the execution period.
Finally,
in contrast to the alternative approaches,
the strategies obtained with our model are static and their computation avoids relying on a computationally intensive dynamic programming approach.
}

The rest of this paper is structured as follows.
In \Cref{sec:models},
we propose a model that incorporates the volume uncertainty in the $\CVaR_\alpha$ equivalent formulation of the return-risk trade-off model of \cite{Almgren2001}.
We establish the relationship between the viability of the market and the uniqueness of a \comadd{convex set of optimal trading strategies.}
\comadd{We also explore avenues of adapting models from the literature to our problem and compare their relative advantages.}
\Cref{sec:numerical_results} provides numerical evidence that our model achieves significantly better mean-$\CVaR_\alpha$ trade-offs than when volume uncertainty is neglected,
\comadd{and it compares our approach against the current alternatives.}
\comadd{We then illustrate how our model can be applied to power trading in Great Britain,
where power wholesalers exchange power future contracts to hedge against the risk associated with the volatility of the power demand.}
Finally,
\Cref{sec:comments,sec:conclusion} discuss the results,
propose avenues for further research on this subject,
and conclude the paper.


%% file: sections/2-models.tex
\section{Model}
\label{sec:models}

In order to fit with the framework described in the introductory examples,
we formulate the optimal execution problem from the point of view of a trader who desires to acquire their position within a fixed time horizon;
\comrefo{this is equivalent to the liquidation problem presented in \citet{Almgren2001} up to a change of sign in the temporary and permanent impacts induced by the trades as presented hereinafter.}

\input{\pathSectionsOptimalTrading/2.1-price_uncertainty.tex}
\input{\pathSectionsOptimalTrading/2.2-price_volume_uncertainty.tex}
\input{\pathSectionsOptimalTrading/2.3-frameworks_comparison.tex}

%% file: sections/2.1-price_uncertainty.tex
\subsection{Optimal execution under price uncertainty}%
\label{sub:mean_variance_framework}

We use the following notation: 
$\ftarget_{T} > 0 $ is the total volume to be traded by time $T$ over $m$ execution periods,
$\rprice_{0}$ is the initial security price,
and~$\tau_{i}$ is the length of the trading period between the two consecutive discrete decision times $t_{i-1}$ and $t_{i}$;
by abuse of language we also use~$\tau_{i}$ in the sense of the $i$-th trading period.
Like \citet{Almgren2001},
we consider that the price dynamics follows an arithmetic random walk,
\comreft{where the price step between decision times $t_{i-1}$ and $t_{i}$ varies according to the distribution of a random variable~$\rpriceshift_{i}$}.
Finally, 
if,
additionally to these assumptions, 
the market temporary and permanent impacts induced by the trades are inserted in the price dynamics,
we obtain that the prices evolve as follows for \comadd{$i \in \dseto{m}$}:

\begin{subequations}
  \label{eq:price_dynamics}
  \begin{align}
    \rprice_{i}          & = \rprice_{i-1} + \rpriceshift_{i} + \tau_{i} g\pa{\frac{\rquant_{i}}{\tau_{i}}}, \label{eq:price_dynamics_permanent_impact}\\
    \tilde{\rprice}_{i}  & = \rprice_{i-1} + h\pa{\frac{\rquant_{i}}{\tau_{i}}}, \label{eq:price_dynamics_temporary_impact}
  \end{align}
\end{subequations}

\noindent
where $\rprice_{i}$ is the security price at decision time~$t_{i}$,
$\rquant_{i}$ is the volume of securities traded during trading period~$\tau_{i}$ \comreft{(bought if $\rquant_{i} > 0$ and sold if $\rquant_{i} < 0$)},
$\tilde{\rprice}_{i}$ is the effective security price for the trades executed during trading period~$\tau_{i}$,
and finally $g$ and $h$ respectively model the permanent and temporary price impact as a function of the average trading rate over the trading interval.
The \emph{liquidation cost} is then a random variable given by~\eqref{eq:mean_variance_trading_cost},
here formulated from a buyer's perspective:

\begin{align}
  \Cost
  \pa{\rquantb} 
  & \mdef \sum_{i=1}^{m} \rquant_{i}\tilde{\rprice}_{i} - \ftarget_{T} \rprice_{0} \nonumber \\
  & =     \sum_{i=1}^{m} \pa{ \rpriceshift_{i} + \tau_{i} g \pa{\frac{\rquant_{i}}{\tau_{i}}}} \pa{\ftarget_{T} - \sum^{i}_{k=1} \rquant_k} + \sum_{i=1}^{m} \rquant_{i} h \pa{\frac{\rquant_{i}}{\tau_{i}}}.
  \label{eq:mean_variance_trading_cost}
\end{align}

\noindent where $\rquantb = \mvec{\rquant_1}{\rquant_m}^{\T}$ is the vector of the trade volumes~$\rquant_{i}$ of period~$\tau_{i}$.
Given the mean-variance framework of \citet{Markowitz1952} and given a risk-aversion parametrised by~$\lambda_{\Var} \geq 0$,
the optimal trading strategy deriving from the model of \citet{Almgren2001} is obtained by solving the following optimisation problem:
\begin{mini!}|l|[2]
  {\rquantb}
  { \expe{\Cost \pa{\rquantb}} + \lambda_{\Var} \vari{\Cost\pa{\rquantb}} \label{eq:mean_variance_objective}}
  {\label{eq:mean_variance_optimisation_problem}}
  {}
  \addConstraint{\1^{\T}\rquantb}{= \ftarget_{T} \label{eq:mean_variance_constraint}}
\end{mini!}
\noindent where~$\1$ is a vector of 1s, 
\ie $\mvec{1}{1}^{\T}$.

\citet[Corollary~1]{Gatheral2012} proved that the continuous formulation of this model is free of price manipulations,
which is tantamount to the market viability,
if the following conditions are satisfied:
\li~$g$ is a linear nondecreasing function,
\ie $g\pa{v}  =  \gamma v$ with $\gamma \geq 0$,
and \lii~the function $f: x \mapsto x h\pa{x}$ is convex.
These conditions are empirically observed as shown in \citet{Almgren2005}.
\comadd{
For the discrete formulation of \citet{Almgren2001},
\citet[Proposition~2]{Huberman2004} provide the following necessary conditions for a market with time-independent price-impact functions to be viable:
\li the linearity of the function~$g$,
and \lii
}

\comadd{
\begin{equation}
  \label{eq:viability_condition_Huberman_Stanzl}
  h \pa{\frac{q}{\tau}} - h \pa{\frac{-q}{\tau}}
  \gtreqless
  {\comadd{g\pa{\frac{q}{\tau}}}},
  \text{ for } q \gtreqless 0,
\end{equation}
}

\noindent
\comadd{
where~$q$ is the volume traded and $\tau$ is the length of each trading period,
\ie $\forall i \in \dseto{m}: \tau_{i} = \tau$;
}
an interpretation of this condition is provided later on.
In their paper, 
\citet{Almgren2001} 
assume linear permanent and temporary impact functions.
We make similar assumptions for our model,
however we allow for the liquidity parameters to vary over time.
Hence, 
\comadd{one has time-dependent price-impact functions.
The permanent and temporary impact functions~$g_{\comadd{i}}$ and~$h_{i}$ related to the trading period~$\tau_{i}$ are noted as follows}
\begin{subequations}
  \begin{align}
    g_{\comadd{i}}(v)  & = \gamma_{\comadd{i}} v \label{eq:linear_permanent_impact},\\
    h_{i}(v)  & = \epsilon_{i} \sign{v} + \eta_{i} v \label{eq:linear_temporary_impact},
  \end{align}
\end{subequations}

\noindent
where~$v$ is the average trading rate over trading period $\tau_{i}$.
This choice of functions is relevant in that parameters~$\epsilon_{i}$ and~$\eta_{i}$ might be considered as the fixed and variable costs of trading.
A reasonable estimate for~$\epsilon_{i}$ is the sum of half the bid-ask spread and the trading fees 
\citep{Almgren2001}.
The parameter~$\eta_{i}$ can be interpreted as the gradient of a linear model for the volumes of orders in the limit order book as a function of the deviation from the best limit order;
\comreft{
  we assume that this parameter is the same when taking either short or long positions.
}%
As a consequence of the temporary impact,
any market participant incurs an additional cost of
\begin{equation}
  \rquant h_{i}\pa{\frac{\rquant}{\tau_{i}}} = \epsilon_{i} \, \abs{\rquant} + \frac{\eta_{i} \rquant^2}{\tau_{i}},
\end{equation}
for trading~$\rquant$ units of the security during the trading period~$\tau_{i}$.

%% file: sections/2.2-price_volume_uncertainty.tex
\subsection{Incorporation of the uncertainty of the volume target}%
\label{sub:mean_CVaR_framework}

In the case where~$\ftarget_{T}$,
the total volume to be traded by time~$T$,
only becomes fully deterministically known during the execution period,
Optimisation Problem~\eqref{eq:mean_variance_optimisation_problem} must be adapted in order to ensure that the trader acquires the right number of units of the security.
For this to happen,
we make the assumption that the volume target is perfectly known at the start of the last trading period~$\tau_m$.
\comadd{
  As of now,
  we denote the volume target by a capital letter,
  \ie $\rtarget_{T}$,
  to emphasise the fact that it is a random variable unlike in \Cref{sub:mean_variance_framework}.}%

In this paper,
we consider that the uncertainty associated with the total volume to be traded is defined as follows:
for a given delivery time $T$,
let $\rtarget_{i}$ be the forecast at time~$t_{i}$ of the total volume to be traded by time~$t_{m}=T$,
\begin{equation}
  \label{eq:demand_forecast_cond_expectation}
  \rtarget_{i} \mdef \cexpe{\rtarget_{T}}{\mset{F}_{i}},
\end{equation}
where $\pa{\mset{F}_{i}}_{i=1}^{m}$ is the filtration of sigma algebras that represent the information available at time~$t_{i}$.
Let us assume that two successive forecasts differ by a continuous random variable $\rupdate_{i}$ for $i<m$.
At decision time~$t_{m-1}$,
the volume target is assumed to be perfectly known.
Hence,
the forecasts of $\rtarget_{T}$ evolve as follows over course of the entire execution period:
\begin{equation}%
  \label{eq:demand_dynamics}
  \begin{cases}
    \rtarget_{i} = \rtarget_{i-1} + \rupdate_{i}, & \text{ if } \comadd{i \in \dseto{m-1}}\\
    \rtarget_{m} = \rtarget_{m-1}, 
  \end{cases}
\end{equation}%
where $\rtarget_{m} = \rtarget_{T}$, the total volume to be traded by time~$T$.

Obviously,
in the presence of volume uncertainty,
Constraint \eqref{eq:mean_variance_constraint} of Optimisation Problem~\eqref{eq:mean_variance_optimisation_problem} cannot be enforced since only a rough estimate of the total volume to be traded is known at time~$t_{0}$,
\ie $\rtarget_{0}$.
Nonetheless,
the model of \citet{Almgren2001} may be useful as a planning tool when deployed in conjunction with a systematic recourse whenever an update on the volume target becomes available.
A trader would then,
at each trading period,
recompute and update their strategy based on the newest available volume forecast.
The main flaw in this approach is that a trader is unable to define a static strategy at time~$t_{0}$ that ensures the satisfaction of Constraint~\eqref{eq:mean_variance_constraint}.
Indeed,
the volumes to be traded at each trading period must constantly be updated.
To get around this issue,
we subsequently propose a new way of defining a trader's strategy.
Furthermore, 
to avoid an explosion in the algorithmic complexity of numerical evaluations of the model,
we want to avoid approaches based on dynamic programming and focus on models that account for the impact of recourse actions via the incorporation of a risk term.

First,
unlike \citet{Almgren2001} who define a trader's strategy as the volumes to be traded in each trading period,
we define a trader's strategy as the proportion $\ystrat_{i}$ of the total volume $\rtarget_{T}$ to acquire over the course of each trading period $\tau_{i}$.
Naturally, 
the following constraint must hold to enforce the distribution of the entire volume target over all the trading periods:
\begin{equation} \label{eq:mean_CVaR_strategy_constraint_sum_to_one}
  \sum^{m}_{i=1} \ystrat_{i} = 1.
\end{equation}
As a consequence,
in the ideal situation where $\rtarget_{T}$ is perfectly predictable,
a trader would acquire $\ystrat_{i} \rtarget_{T}$ units of the security during trading period $\tau_{i}$.
Obviously, 
this definition must be adapted in the situation where $\rtarget_{T}$ is uncertain as a trader should take recourse in order to satisfy Constraint~\eqref{eq:mean_variance_constraint}.
As a starting point,
we assume that a trader respects their initial strategy by trading during each trading period~$\tau_{i}$ the planned proportion~$\ystrat_{i}$ of~$\rtarget_{i-1}$,
\ie the best estimate of the volume target $\rtarget_{T}$ available.
This solves only partially the problem engendered by the forecast updates,
since past decisions cannot be modified \emph{a posteriori},
which means that $\sum_{i=1}^{m} \ystrat_{i}\rtarget_{i-1} \neq \rtarget_{T}$.
Indeed,
if we denote by $\rupdate_{k}$,
\comreft{$ k < m $},
the \emph{forecast update} reported at the decision time~$t_{k}$,
then the \emph{forecast error} related to $\rupdate_{k}$ due to the past decisions amounts to $\rupdate_{k}^{\varepsilon} \mdef \rupdate_{k} \sum^{k}_{r=1} \ystrat_{r}$.
This corresponds to the additional volume,
which may be positive or negative,
that needs to be traded over the remaining trading periods \comadd{in order to match with $\rtarget_{T}$}.
We predicate that when an update on the volume target occurs,
the trader adjusts their strategy by redistributing the forecast error over the remaining trading periods according to a fixed distribution independent of the decision variables.
Considering these corrections upfront allows to partially account for future recourse actions without having to compute a costly dynamic programming simulation.
In our approach we redistribute this additional volume over the future trading periods according to fixed proportions.
We treat these fixed proportions as model parameters and leave the learning of optimal parameter values to future work.
Let~$\beta_{k,i}$ denote the proportion of the forecast error~$\rupdate_{k}^{\varepsilon}$ to correct for at trading period~$\tau_{i}$.
Evidently,
$\beta_{k,i} = 0 $ if $i\leq k$ and $ \sum^{m}_{i=k+1} \beta_{k,i} = 1$ for all $k < m$.
\comadd{We will refer to $\betab \mdef \pa{\beta_{k,i}} \in \R^{m-1 \times m}$ as the redistribution matrix.}

In the case where the volume target is uncertain,
we define the cost of a trading strategy
as the difference between
\li the trading cost incurred at the end of the execution period by following the initial trading strategy under the consideration of the rough model on their recourse determined by the redistribution matrix $\betab$
with \lii the trading cost ideally obtained in an infinitely liquid market where the entire position $\rtarget_{T}$ is traded at the start of the execution period.
With these considerations,
a trader's trading cost related to a strategy $\ystratb = \mvec{\ystrat_{1}}{\ystrat_{m}}^{\T}$ is the following random variable:
\begin{align}
  \Cost
  \pa{\ystratb}
  &
  \mdef
  \sum_{i=1}^{m} 
\rquant_{i}\pa{\ystratb}
\tilde{\rprice}_{i}
- \rtarget_{T} \rprice_{0}
  \\
  &
  =
  \sum_{i=1}^{m} \pac{ \pa{ \rpriceshift_{i} + \tau_{i} g_{\comadd{i}}\pa{\frac{\rquant_{i}\pa{\ystratb}}{\tau_{i}}} }  \pa{\rtarget_{0} + \sum_{k=1}^{m} \rupdate_{k}-\sum^{i}_{k=1} \rquant_{k}\pa{\ystratb}} }
  + \sum^{m}_{i=1} \rquant_{i}\pa{\ystratb} h_{i}\pa{\frac{\rquant_{i}\pa{\ystratb}}{\tau_{i}}},
  \label{eq:mean_CVaR_trading_cost}
\end{align}
where $\rquant_{i}\pa{\ystratb}$ is a random variable representing the volume to be traded during trading period~$\tau_{i}$:
\begin{equation}
  \label{eq:relation_volume_proportion}
  \rquant_{i}\pa{\ystratb} =
  \ystrat_{i} \rtarget_{i-1}
  + \sum^{i-1}_{k=1} \beta_{k,i} \, \rupdate^{\varepsilon}_{k}
  =
  \ystrat_{i} \rtarget_{0} + \sum^{i-1}_{k=1} \rupdate_{k} \pa{ \ystrat_{i} + \beta_{k,i} \sum^{k}_{r=1} \ystrat_{r}}.
\end{equation}
The \comadd{number of positions} $\rquant_{i}\pa{\ystratb}$ to trade at trading period $\tau_{i}$ is composed of
\li the proportion $\ystrat_{i}$ of the best volume estimate $\rtarget_{i-1}$,
and \lii the position adjustments related to the previous volume forecast updates.
We now have that if $\forall i \in \dseto{m}$, 
$\rquant_{i}\pa{\ystratb}$ is defined as in \eqref{eq:relation_volume_proportion},
then $\1^{\T}\rquantb\pa{\ystratb} = \rtarget_{T}$,
where $\rquantb\pa{\ystratb} \mdef \mvec{\rquant_1\pa{\ystratb}}{\rquant_{m}\pa{\ystratb}}^{\T}$. 
From~\eqref{eq:relation_volume_proportion},
we observe that,
given any outcome $\omega$ of the sample space $\Omega$ and thus any realisation of the forecast updates,
\ie $\rupdateb \pa{\omega} \mdef \mvec{\rupdate_1\pa{\omega}}{\rupdate_{m\comadd{-1}}\pa{\omega}}^{\T}$,
the realisation of the volumes to trade $\rquantb \pgivenomega{\ystratb} \mdef$ $\mvec{(\rquant_1\pa{\ystratb})\pa{\omega}}{(\rquant_{m}\pa{\ystratb})\pa{\omega}}^{\T}$
can be expressed as a linear combination of the decision variables $\ystratb$.
Formally this mean that $\forall \omega \in \Omega$,
one can find a (lower triangular) matrix $\Lb\pa{\omega}$ such that
\begin{equation}
  \label{eq:matrix_relation_volume_proportion}
  \rquantb \pgivenomega{\ystratb} = \Lb\pa{\omega} \ystratb.
\end{equation}
Moreover,
if $\rtarget_{0} \neq 0$,
$\Lb\pa{\omega}$ is non-singular as all elements on the diagonal are different from zero, \ie $\forall i: \Lb_{ii}\pa{\omega} = \rtarget_{0}$.
In the rest of the paper,
$\Cost \pgivenomega{\ystratb}$ is shorthand for $\pa{\Cost\pa{\ystratb}}\pa{\omega}$,
the trading cost of executing strategy $\ystratb$ given the outcome $\omega$.

Finally,
we consider that if a trader is guaranteed to not pay excessive prices in adverse times,
the variance of their trading cost is not of foremost importance.
We thus consider that a risk-averse trader is more interested in minimising their expected trading cost conditional to a quantile of worst case scenarios rather than minimising the variance of their trading cost over all scenarios.
We therefore quantify the risk of a trader's strategy with the \emph{$\alpha$-Conditional Value-at-Risk} ($\CVaR_{\alpha}$) risk measure:
\begin{equation}
  \label{eq:CVaR_definition}
  \cvar{\alpha}{\Cost\pa{\ystratb}} = \frac{1}{\beta_{\ystratb}} \int_{\Omega_{\ystratb}} \Cost \pgivenomega{\ystratb} \dP\pa{\omega},
\end{equation}
where $\beta_{\ystratb} = \proba{\Cost\pa{\ystratb} \geq \varr{\alpha}{\Cost\pa{\ystratb}}}$,
$\Omega_{\ystratb} = \pab{\omega \in \Omega : \Cost \pgivenomega{\ystratb} \geq \varr{\alpha}{\Cost\pa{\ystratb}}}$,
and
\begin{equation}
  \label{eq:VaR_definition}
  \varr{\alpha}{\Cost\pa{\ystratb}} = \min \setc{\gamma \in \R}{F_{\Cost\pa{\ystratb}}(\gamma) \geq 1-\alpha},
\end{equation}

\noindent
where

\begin{equation}
  \label{eq:trading_cost_cdf}
  F_{\Cost\pa{\ystratb}}(\gamma) = \Prob \pac{\Cost\pa{\ystratb} \leq \gamma}
\end{equation}

\noindent
is the cumulative distribution function of $\Cost\pa{\ystratb}$.
The $\alpha$-Conditional Value-at-Risk,
is a coherent risk measure 
\citep{Artzner1999,Rockafellar2002}
that focuses on the \comadd{proportion $\alpha$} of extreme costs and can be interpreted as the expectation of the costs conditional on them exceeding the threshold $\varr{\alpha}{\Cost\pa{\ystratb}}$.
Given a risk-aversion parameter $\lambda_{\CVaR} \in \cset{0}{1}$,
a trader thus tries to minimise the $\meanCVaR_{\alpha}$ trade-off of the total trading cost:

\begin{mini!}|l|[2]                       
  {\ystratb}  
  {\varphi_{\comadd{\alpha}}^{\lambda_{\CVaR}}\pa{\ystratb} \mdef \pa{1 - \lambda_{\CVaR}}\expe{\Cost\pa{\ystratb}} + \lambda_{\CVaR} \cvar{\alpha}{\Cost\pa{\ystratb}} \label{eq:mean_CVaR_objective}}   
  {\label{eq:mean_CVaR_optimisation_problem}}             
  {}                                
  \addConstraint{\1^{\T}\ystratb }{= 1. \label{eq:mean_CVaR_constraint}}    
\end{mini!}
Note that in Optimisation Problem~\eqref{eq:mean_CVaR_optimisation_problem},
a trader's recourse is estimated upfront rather than being simulated via dynamic programming to avoid the curse of dimensionality that plagues stochastic optimisation.
Nevertheless,
in contrast to the model of \citet{Almgren2001},
Model~\eqref{eq:mean_CVaR_optimisation_problem} is well-defined since it adapts to the total \comadd{volume} variability and is thus guaranteed to satisfy the constraint that the total traded volume equals $\rtarget_{T}$.

\subsubsection{A necessary condition for market viability}%
\label{ssub:market_viability_necessary_condition}

\comadd{
  As previously mentioned,
  \citet{Huberman2004} provide the necessary condition \eqref{eq:viability_condition_Huberman_Stanzl} for a market to be viable if the trading periods are homogeneous,
  \ie a market with $g_i = g$, $h_i = h$ and $\tau_{i} = \tau$,  for every trading period $\tau_{i}, i \in \dseto{m}$.
  In the case of linear impact functions,
  \ie \Cref{eq:linear_permanent_impact,eq:linear_temporary_impact},
  Condition~\eqref{eq:viability_condition_Huberman_Stanzl} then becomes
  \begin{equation}
    \label{eq:mean_CVaR_viability_condition_time_independent}
    \pa{ \epsilon \sign{\frac{q}{\tau}} +  \eta \pa{ \frac{q}{\tau} } } 
    -
    \pa{ \epsilon \sign{\frac{-q}{\tau}} +  \eta \pa{ \frac{-q}{\tau} } } 
    \gtreqless
    \gamma q,
    \text{ for }
    q \gtreqless 0.
  \end{equation}
}%
\noindent
This condition intuitively means that the temporary cost incurred by buying and reselling any same amount $q$ of shares on different trading periods should always be greater than the cost saving opportunity engendered by the price shift caused by the permanent impact.
For any vector $\epsilonb \mdef \mvec{\epsilon_{1}}{\epsilon_{m}}^{\T}$ with $\forall k: \epsilon_{k} \geq 0$,
Condition~\eqref{eq:mean_CVaR_viability_condition_time_independent} is equivalent to $\eta > \frac{1}{2}\gamma\tau$,
which is the sufficient condition for having a unique optimal trading strategy in the model of \citet{Almgren2001}.

\comadd{
  With the rationale that the market should not allow any price manipulation,
  which we define as a round-trip strategy with negative expected cost,
  Condition~\eqref{eq:mean_CVaR_viability_condition_time_independent} can be generalised in the more general framework where the impact functions $g_{i}$ and $h_{i}$ depend on the trading period $\tau_{i}$.
}

\begin{definition}[Round-trip strategy]\label{def:round_trip_strategy}%
  A \emph{round-trip strategy} $\rquantb$ is a strategy where the sum of all the trades executed during the execution period equates to zero,
  \ie $\1^{\T}\rquantb=0$.
\end{definition}
\noindent Note that a round-trip strategy is defined in terms of the volumes to trade during each trading period rather than the proportions $\ystratb$.
Indeed,
in the case where the total volume to trade $\rtarget_{T}$ is certain and equals zero,
the only round-trip strategy based on the proportions of $\rtarget_{T}$ would be equivalent to a no trade;
hence defining a round-trip strategy in terms of $\rquantb$ offers more flexibility.
Based on \Cref{def:round_trip_strategy} and on the price manipulation definition from \citet[Definition 1]{Huberman2004},
we define a \emph{price manipulation} as follows:
\begin{definition}[Price manipulation]\label{def:price_manipulation}%
  A (risk neutral) \emph{price manipulation} is a round-trip strategy $\rquantb$ with a strictly negative expected cost,
  \ie
  \begin{equation}\label{eq:price_manipulation}
    \expe{\Cost(\rquantb)} < 0.
  \end{equation}	
\end{definition}

\begin{lemma}\label{lemma:mean_CVaR_market_viability}%
  If a market is free of price manipulation,
  then the matrix $\Mb \mdef \Eb - \Gammab$ is positive semi-definite,
  \ie $\Mb \succeq 0$, where
  \begin{equation}
    \Eb
    = 
    \begin{bmatrix}
      \frac{2\eta_{1}}{\tau_{1}} + \frac{2\eta_{m}}{\tau_{m}} & \frac{2\eta_{m}}{\tau_{m}} & \cdots                     & \frac{2\eta_{m}}{\tau_{m}} \\
      \frac{2\eta_{m}}{\tau_{m}}                              & \ddots                     & \ddots                     & \vdots \\
      \vdots                                                  & \ddots                     & \ddots                     & \frac{2\eta_{m}}{\tau_{m}} \\
      \frac{2\eta_{m}}{\tau_{m}}                              &  \cdots                    & \frac{2\eta_{m}}{\tau_{m}} & \frac{2\eta_{m-1}}{\tau_{m-1}} + \frac{2\eta_{m}}{\tau_{m}}
    \end{bmatrix}
    \text{, and } \,
    \Gammab
    = 
    \begin{bmatrix}
      2\gamma_{1}   & \gamma_{2}  &         & \gamma_{i}  &        & \gamma_{m-1}  \\
      \gamma_{2}    & 2\gamma_{2} &         & \vdots      &        & \vdots        \\
		    &             & \ddots  & \vdots      &        & \vdots        \\
      \gamma_{i}    & \cdots      & \cdots  & 2\gamma_{i} &        & \vdots        \\
		    &             &         &             & \ddots & \vdots        \\
      \gamma_{m-1}  & \cdots      & \cdots   & \cdots     & \cdots & 2\gamma_{m-1} \\         
    \end{bmatrix}.
  \end{equation}
\end{lemma}
\begin{proof}
  To prove \Cref{lemma:mean_CVaR_market_viability},
  we show that if matrix $\Mb$ is not positive semi-definite,
  then we can find a price manipulation.
  Given an outcome $\omega \in \Omega$,
  the trading cost of a round-trip strategy $\rquantb$ is given by

  \begin{equation}
    \label{eq:trading_cost_realisation_round_trip}
    \Cost
    \pgivenomega{\rquantb}
    =
    \sum_{i=1}^{m} \pac{\pa{ \rpriceshift_{i}\pa{\omega} + \gamma_{i} \rquant_{i}}  \pa{-\sum^{i}_{k=1} \rquant_{k}}}
    + \sum^{m}_{i=1} \pa{ \epsilon_{i} \abs{\rquant_{i}} + \frac{\eta_{i}\rquant_{i}^{2}}{\tau_{i}} }.
  \end{equation}

  \noindent
  As $\rquantb$ defines a round-trip strategy,
  the trading cost can be expressed in terms of the first $m-1$ components of $\rquantb$ by using the fact that $\sum_{i=1}^{m}\rquant_{i} = 0 \Rightarrow \rquant_{m} = -\sum_{i=1}^{m-1} \rquant_{i}$:

  \begin{multline}
    \label{eq:trading_cost_realisation_round_trip_with_constraint}
    \Cost
    \pgivenomega{\rquantb}
    =
    \sum_{i=1}^{m-1} \pac{\pa{ \rpriceshift_{i}\pa{\omega} + \gamma_{i} \rquant_{i}}  \pa{-\sum^{i}_{k=1} \rquant_{k}}}
    + \sum^{m-1}_{i=1} \pa{ \epsilon_{i} \abs{\rquant_{i}} + \frac{\eta_{i}\rquant_{i}^{2}}{\tau_{i}}} \\
    + \pa{ \epsilon_{m} \abs{\sum_{i=1}^{m-1} \rquant_{i}} + \frac{\eta_{m}\pa{\sum_{i=1}^{m-1} \rquant_{i}}^{2}}{\tau_{m}}}.
  \end{multline}

  \noindent
  This trading cost can be split in two parts:
  \li $Lin \pgivenomega{\rquantb}$ which grows linearly with $\rquantb$ and depends on the uncertainty outcome $\omega$,
  and \lii a quadratic term in $\rquantb$ independent of $\omega$:
  \begin{align}
    \label{eq:decomposition_parts}
    \Cost
    \pgivenomega{\rquantb}
    & =
    Lin \pgivenomega{\rquantb}
    + \sum^{m-1}_{i=1} \pa{\frac{\eta_{i}\rquant_{i}^{2}}{\tau_{i}}}
    + \frac{\eta_{m}\pa{\sum_{i=1}^{m-1} \rquant_{i}}^{2}}{\tau_{m}}
    - \sum_{i=1}^{m-1} \pa{\gamma_{i} \rquant_{i} \pa{\sum^{i}_{k=1} \rquant_{k}}}\\
    & = 
    Lin \pgivenomega{\rquantb}
    +
    \frac{1}{2}\rquantb_{[1:m-1]}^T \, \Mb \,\rquantb_{[1:m-1]},
  \end{align}
  where $\rquantb_{[1:m-1]}$ is the vector of the first $m-1$ components of $\rquantb$.
  If matrix $\Mb$ is not positive semi-definite,
  there exists a direction $\db \in \R^{m-1}$ along which the quadratic form is concave for every $\omega \in \Omega$ as $\Mb$ is independent of $\omega$;
  any normalised eigenvector associated to one of the negative eigenvalues of $\Mb$ is such a direction.
  In the following, 
  let $\zeta^{-}$ be a negative eigenvalue of $\Mb$ and $\vb$ an associated normalised eigenvector. 
  As a consequence,
  there exists a $\kappa \in \R$,
  such that the round-trip strategy $\tilde{\rquantb} \mdef \pac{ \kappa v_1, \dots, \kappa v_{m-1}, - \kappa \pa{\1^{\T}\vb} }^{\T}$ is a price manipulation.
  Indeed,
  one can find a $\kappa \in \R$,
  such that $\forall \omega \in \Omega: \, \Cost \pgivenomega{\tilde{\rquantb}} < 0$, 
  which implies that $\tilde{\rquantb}$ is a price manipulation, 
  \ie $\expe{\Cost\pa{\tilde{\rquantb}}} < 0$.
  The existence of such a $\kappa$ is justified by the fact that when $\kappa \to \infty$,
  and thus when $\norm{2}{\tilde{\rquantb}} \to \infty$, 
  the trading cost is dominated by its quadratic part,
  \ie $\Cost\pgivenomega{\tilde{\rquantb}} \simeq \, \frac{1}{2} \tilde{\rquantb}_{[1:m-1]}^{\T} \, \Mb \,\tilde{\rquantb}_{[1:m-1]} = \frac{1}{2}\kappa^{2}\zeta^{-} < 0$.
\end{proof}

\subsubsection{Set of optimal strategies}%
\label{ssub:optimal_strategy_uniqueness}

Analogously \comadd{than for} Model~\eqref{eq:mean_variance_optimisation_problem},
we can prove that the objective function $\varphi_{\comadd{\alpha}}^{\lambda_{\CVaR}}$ of Optimisation Problem~\eqref{eq:mean_CVaR_optimisation_problem} is convex on its feasible set $\mset{Y} \mdef \setc{\ystratb \in \R^{m}}{\1^{\T}\ystratb = 1}$ \comadd{ if the market excludes any price manipulation}.
The convexity of the objective function on $\mset{Y}$ for any value of $\lambda_{\CVaR} \in \cset{0}{1}$ is equivalent to the simultaneous convexity of both functions 
$f_1 : \R^{m} \to \R; \ystratb \mapsto \expe{\Cost\pa{\ystratb}}$ and $f_2: \R^{m} \to \R; \ystratb \mapsto \cvar{\alpha}{\Cost\pa{\ystratb}}$ on $\mset{Y}$.

\begin{lemma}
  \label{lemma:mean_CVaR_convexity_expectation}%
  If the permanent and temporary impact functions are given by \eqref{eq:linear_permanent_impact}-\eqref{eq:linear_temporary_impact} and if \comadd{$\Mb \succeq 0$} \comadd{(\resp $\Mb \succ 0$)}, 
  then function $f_1:\R^{m} \to \R; \ystratb \mapsto \expe{\Cost\pa{\ystratb}}$ is \comadd{(\resp strictly)} convex on $\mset{Y}$.
\end{lemma}
\begin{proof}
  By definition,  we have $\expe{\Cost\pa{\ystratb}} = \int_{\Omega} \Cost \pgivenomega{\ystratb} \dP\pa{\omega}$.
  Hence,
  if for any outcome $\omega \in \Omega$,
  function $\Cost \pdotomega : \R^{m} \to \R; \ystratb \mapsto \Cost \pgivenomega{\ystratb}$ is \comadd{(\resp strictly)} convex in $\ystratb$,
  then the result of \Cref{lemma:mean_CVaR_convexity_expectation} holds as the expectation of $\Cost\pa{\ystratb}$ can be seen as a non-negative weighted sum of the $\Cost \pgivenomega{\ystratb}$.
  Given~\eqref{eq:mean_CVaR_trading_cost}, 
  $\Cost \pgivenomega{\ystratb}$ is written as follows:

  \begin{multline}
    \Cost
    \pgivenomega{\ystratb}
    =  
    \sum_{i=1}^{m} \pac{\pa{ \rpriceshift_{i}\pa{\omega} + \tau_{i} g_{\comadd{i}}\pa{\frac{\rquant_{i} \pgivenomega{\ystratb}}{\tau_{i}}}}  \pa{\rtarget_{0} +
    \sum_{k=1}^{m} \rupdate_{k}\pa{\omega}-\sum^{i}_{k=1} \rquant_{k}\pgivenomega{\ystratb}}} \\
    + \sum^{m}_{i=1} \rquant_{i} \pgivenomega{\ystratb} h_{i}\pa{\frac{\rquant_{i}\pgivenomega{\ystratb}}{\tau_{i}}}.
  \end{multline}

  \noindent
  Since for any $\omega \in \Omega$,
  there exists a linear transformation between $\ystratb$ and $\rquantb \pgivenomega{\ystratb}$ as shown in~\eqref{eq:matrix_relation_volume_proportion},
  proving the convexity of $\Cost \pdotomega $ in $\ystratb$ is equivalent to proving the convexity of $\widetilde{\Cost} \pdotomega$ in $\rquantb \pgivenomega{\ystratb}$ \citep{Boyd2004},
  where,
  using \eqref{eq:linear_permanent_impact}-\eqref{eq:linear_temporary_impact},
  $\widetilde{\Cost} \pdotomega$ is given by

  {
    \footnotesize
    \begin{align}
      \widetilde{\Cost}
      \pgivenomega{ \rquantb \pgivenomega{\ystratb} }
      = \;& 
      \sum_{i=1}^{m} \pac{ \pa{ \rpriceshift_{i}\pa{\omega} + \gamma_{\comadd{i}} \rquant_{i}\pgivenomega{\ystratb}} 
      \pa{\rtarget_{0} + \sum_{k=1}^{m} \rupdate_{k}\pa{\omega}-\sum^{i}_{k=1} \rquant_{k}\pgivenomega{\ystratb}}  }
      + \sum^{m}_{i=1} \pac{\epsilon_{i} \abs{\rquant_{i}\pgivenomega{\ystratb}} + \frac{\eta_{i}\rquant_{i}^{2}\pgivenomega{\ystratb}}{\tau_{i}}} \nonumber \\
      = \;&%
      \sum_{i=1}^{m} \rpriceshift_{i}\pa{\omega}
      \pa{\rtarget_{T}\pa{\omega} -  \sum^{i}_{k=1} \rquant_{k}\pgivenomega{\ystratb}}
      + \sum_{i=1}^{m} \gamma_{\comadd{i}} \rquant_{i} \pgivenomega{\ystratb} \rtarget_{T}\pa{\omega}
      + \sum^{m}_{i=1} \epsilon_{i} \abs{ \rquant_{i} \pgivenomega{\ystratb} } 
      \nonumber \\ 
	  &
	  -  \sum_{i=1}^{m} \gamma_{\comadd{i}} \rquant_{i} \pgivenomega{\ystratb} \pa{\sum^{i}_{k=1} \rquant_{k} \pgivenomega{\ystratb}}
	  + \sum^{m}_{i=1} \frac{\eta_{i}\rquant_{i}^{2} \pgivenomega{\ystratb} }{\tau_{i}},
	  \label{eq:mean_CVaR_trading_cost_realisation_aux}%
    \end{align}%
  }%
  where $\rtarget_{T} \pa{\omega} \mdef \rtarget_{0} + \sum_{k=1}^{m} \rupdate_{k} \pa{\omega}$.
  \comadd{
    It is straightforward that the first three terms of \eqref{eq:mean_CVaR_trading_cost_realisation_aux} are convex in $\rquantb \pgivenomega{\ystratb}$,
    the last two can be combined  in a similar manner as in the proof of \Cref{lemma:mean_CVaR_market_viability},
    \ie
    \begin{multline}
      \label{eq:reformulation_quadratic_term}
      -  \sum_{i=1}^{m} \gamma_{\comadd{i}} \rquant_{i} \pgivenomega{\ystratb} \pa{\sum^{i}_{k=1} \rquant_{k} \pgivenomega{\ystratb} }
      + \sum^{m}_{i=1} \frac{\eta_{i}\rquant_{i}^{2} \pgivenomega{\ystratb}}{\tau_{i}}
      = 
      \frac{1}{2}\rquantb_{[1:m-1]}^T \, \Mb \,\rquantb_{[1:m-1]} \\
      - \gamma_{m} \pa{ \rtarget_{T}\pa{\omega} -  \sum_{i=1}^{m-1} \rquant_{i} \pgivenomega{\ystratb} } \rtarget_{T}\pa{\omega} 
      + \frac{\eta_{m}}{\tau_{m}}   \rtarget_{T}\pa{\omega} \pa{ \rtarget_{T}\pa{\omega} - 2\sum_{i=1}^{m-1} \rquant_{i} \pgivenomega{\ystratb} }
      ,
    \end{multline}
    where $\rquantb_{[1:m-1]} \mdef \mvec{\rquant_{1} \pgivenomega{\ystratb} }{\rquant_{m-1} \pgivenomega{\ystratb} }^{\T}$.
    Hence the quadratic part of the trading costs is (\resp strictly) convex if $\Mb \succeq 0$ (\resp $\Mb \succ 0$);
    this terminates the proof.
  }
\end{proof}

\begin{lemma}
  \label{lemma:mean_CVaR_convexity_CVaR}%
  If the permanent and temporary impact functions are given by \eqref{eq:linear_permanent_impact}-\eqref{eq:linear_temporary_impact}
  and if \comadd{$\Mb \succeq 0$} \comadd{(\resp $\Mb \succ 0$)}, 
  then function $f_2: \R^{m} \to \R; \ystratb \mapsto \cvar{\alpha}{\Cost\pa{\ystratb}}$ is \comadd{(\resp strictly)} convex on $\mset{Y}$.
\end{lemma}
\begin{proof}
  Any real valued coherent risk measure $\varrho: \R^{m} \to \R; \ystratb \mapsto \varrho \pac{ \Cost\pa{\ystratb} }$ as defined in \citet{Artzner1999},
  can be expressed in its dual representation,
  \ie
  \begin{equation}
    \label{eq:coherent_risk_measure_dual}
    \varrho \pac{\Cost\pa{\ystratb}} = \sup_{\mu \in \mset{D}} \expem{\mu}{\Cost\pa{\ystratb}},
  \end{equation}
  where $\mset{D}$ is a convex subset of probability measures on $\Omega$ \citep[Theorems 6.4 and 6.6]{Shapiro2009a}.
  As a consequence, 
  as $\CVaR_{\alpha}$ is a coherent risk measure \citep[Corollary 12]{Rockafellar2002},
  the desired result is shown by combining the (\resp strict) convexity of $\Cost\pdotomega$ for any $\omega$ with the fact that a function defined as the pointwise supremum of (\resp strictly) convex functions is (\resp strictly) convex \citep{Boyd2004}.
\end{proof}

\begin{theorem}\label{theorem:mean_CVaR_convexity_model}%
  If the permanent and temporary impact functions are given by \eqref{eq:linear_permanent_impact}-\eqref{eq:linear_temporary_impact},
  if $\lambda_{\CVaR} \in \cset{0}{1}$ and if \comadd{$\Mb \succeq 0$} \comadd{(\resp $\Mb \succ 0$)}, 
  then the objective function of Optimisation Problem~\eqref{eq:mean_CVaR_optimisation_problem} is \comadd{(\resp strictly)} convex on its feasible set $\mset{Y}$.
\end{theorem}
\begin{proof}
  The proof is straightforward based on \Cref{lemma:mean_CVaR_convexity_expectation,lemma:mean_CVaR_convexity_CVaR}.
\end{proof}

Similarly to Model~\eqref{eq:mean_variance_optimisation_problem}, 
the convexity and the non-emptiness of the feasible set~$\mset{Y}$ 
and \Cref{theorem:mean_CVaR_convexity_model} imply, 
under the conditions that $\lambda_{\CVaR} \in \cset{0}{1}$ and that \comadd{$\Mb \succeq 0$,
the existence and uniqueness of a convex set $\mset{Y}^{\star} \subseteq \mset{Y}$ of optimal execution strategies for} the model defined in~\eqref{eq:mean_CVaR_optimisation_problem}.
\comadd{If $\Mb \succ 0$,
  the set of optimal strategies is a singleton, 
  \ie $\mset{Y}^{\star} \mdef \pab{\ystratb^{\star}}$.
}

\comadd{
  In other words,
  the necessary condition $\Mb \succeq 0$ for the absence of price manipulation,
  and thus for having a viable market where no trader has interest in impacting the price dynamics with artificial trades,
  is tantamount to the convexity of Optimisation Problem~\eqref{eq:mean_CVaR_optimisation_problem},
  and thus to a convex set of optimal strategies for a trader considering both price and volume uncertainty.
}

%% file: sections/2.3-frameworks_comparison.tex
\subsection{\comreft{Comparison with other approaches}}%
\label{sub:frameworks_comparison_theory}

\paragraph{Mean-variance with recourse.}%
\label{par:mean_variance_with_recourse}

In the case where the volume target is uncertain,
the model of \citet{Almgren2001} can be deployed in conjunction with a systematic recourse whenever a forecast update on the volume target becomes available.
We will show that such a strategy can be reproduced in our model with an appropriate choice of $\ystratb$ and $\betab$.
Similarly to Model~\eqref{eq:mean_CVaR_optimisation_problem},
a trading strategy~$\nb$ in Model~\eqref{eq:mean_variance_optimisation_problem} can be equivalently expressed as the proportions~$y_{i}$ of the fixed volume target~$\ftarget_{T}$ to trade during each trading period~$\tau_{i}$, 
$i \in \dseto{m}$.

\begin{lemma}
  \label{lemma:mean_variance_strategy_independent_of_volume}%
  If $\rpriceshift_{i}$,
  $i \in \dseto{m}$,
  are draws from independent random variables each with zero mean,
  \ie $\expe{\xi_i}= 0$,
  if the permanent and temporary impact functions are given by \eqref{eq:linear_permanent_impact}-\eqref{eq:linear_temporary_impact},
  and if the temporary impact parameter $\epsilon_{i}$ is constant over all
  (\ie independent of the)
  trading periods,
  \ie $\forall i \in \dseto{m}: \epsilon_{i} = \epsilon$,
  then the optimal strategy $\ystratb^{\star}$ of Optimisation Problem~\eqref{eq:mean_variance_optimisation_problem} expressed in terms of the proportions of $\ftarget_{T}$ is independent of $\ftarget_{T}$.
\end{lemma}
\begin{proof}
  Based on \eqref{eq:mean_variance_trading_cost},
  the expectation and variance of the trading cost are given by

  \begin{align*}
    \expe{ \Cost \pa{ \rquantb } }
    & =  
    \sum_{i=1}^{m} \gamma_{i} \rquant_{i} \pa{ \ftarget_{T} - \sum^{i}_{k=1} \rquant_{k} } 
    + \sum_{i=1}^{m} \pa{ \epsilon \abs{\rquant_{i}} + \frac{\eta_{i} \rquant_{i}^{2}}{\tau_{i}} }\\
    & =
    \sum_{i=1}^{m} \gamma_{i} \ystrat_{i} \pa{ 1 - \sum^{i}_{k=1} \ystrat_{k} } \ftarget_{T}^{2} 
    +  \sum_{i=1}^{m} \pa{ \epsilon \abs{y_{i} \ftarget_{T} }
    + \frac{\eta_{i} }{\tau_{i}} \pa{ \ystrat_{i} \ftarget_{T} }^2 }, \\
    \vari{\Cost(\rquantb)} & =
    \sum_{i=1}^{m} \vari{\rpriceshift_{i}} \pa{ \ftarget_{T}
    - \sum^{i}_{k=1} \rquant_{k} }^2
    =
    \sum_{i=1}^{m} \vari{\rpriceshift_{i}} \pa{ 1 - \sum^{i}_{k=1} \ystrat_k }^2\ftarget_{T}^2.
  \end{align*}
  Moreover,
  if we assume \wlogg that $\ftarget_{T} > 0$,
  then the optimal volumes $\rquantb^\star$ of Optimisation Problem~\eqref{eq:mean_variance_optimisation_problem} are positive \citep{Almgren2001}.
  Therefore,
  adding the non-negativeness constraint on the decision variables $\rquantb$ to Optimisation Problem~\eqref{eq:mean_variance_optimisation_problem} does not alter its optimal solution.
  Hence,
  one can replace $\sum_{i=1}^{m} \epsilon \abs{\rquant_{i}} $ by $ \sum_{i=1}^{m} \epsilon \rquant_{i} = \epsilon \ftarget_{T}$.
  As a consequence,
  all the terms in the objective function of Optimisation Problem~\eqref{eq:mean_variance_optimisation_problem} that depend on the decision variables $\ystratb$ are multiplied by $\ftarget_{T}^{2}$, 
  which means that $\ftarget_{T}$ has no impact on the optimal solution $\ystratb^{\star}$.
\end{proof}

\noindent
\Cref{lemma:mean_variance_strategy_independent_of_volume} implies that under price and volume uncertainty,
the strategy of a trader consisting in deploying the model of \citet{Almgren2001} in conjunction with a systematic recourse whenever an update on the volume target becomes available is time consistent.
More precisely, 
given the optimal strategy $\ystratb^{\star,t_0}$ of~\eqref{eq:mean_variance_optimisation_problem} computed at time~$t_0$, 
if one recomputes the optimal strategy $\ystratb^{\star,t_{i-1}}$ of~\eqref{eq:mean_variance_optimisation_problem} at a subsequent decision time~$t_{i-1}$,
$ i \in \dset{2}{m}$,
it would simply be the continuation from time~$t_{i-1}$ to $t_m$ of $\ystratb^{\star,t_0}$,
\ie 
\begin{equation}
  \ystratb^{\star,t_{i-1}}
  \mdef 
  \mvec{\ystrat_{i}^{\star,t_{i-1}}}{\ystrat_{m}^{\star,t_{i-1}}}^{\T}
  =
  \pa{ \frac{1}{\sum_{r=i}^{m}{\ystrat^{\star,t_0}_{r}}} } \cdot \ystratb^{\star,t_0}_{[i:m]},
\end{equation}
where $\ystratb^{\star,t_0}_{[i:m]} \mdef \mvec{\ystrat_{i}^{\star,t_{0}}}{\ystrat_{m}^{\star,t_{0}}}^{\T}$.
Note that the rescaled proportions $\ystratb^{\star,t_{i-1}}$,
$i \in \dset{2}{m}$,
are independent of past realisations (price moves and forecast updates) and can consequently be computed beforehand at the start of the execution period,
\ie at time~$t_{0}$.
If $\rquantb^{rec} \mdef  \mvec{\rquant_{1}^{rec}}{\rquant_{m}^{rec}}^{\T}$ denotes the traded volumes of a trader who updates their strategy by solving~\eqref{eq:mean_variance_optimisation_problem} whenever an update on the volume target occurs,
then the volume $\rquant_{i}^{rec}$ traded on trading period $\tau_{i}$,
$i \in \dseto{m}$,
is equal to the proportion $\ystrat^{\star,t_{i-1}}_{i}$ of the best volume target estimate $\rtarget_{i-1}$ minus the volumes already traded,
\ie $ \rquant_{i}^{rec} = \ystrat^{\star,t_{i-1}}_{i} \cdot \pa{\rtarget_{i-1} - \sum_{k=1}^{i-1} \rquant_{k}^{rec}}$.

\begin{lemma}\label{lemma:mean_CVaR_reproducibility_mean_variance_with_recourse}%
  If $\rpriceshift_{i}$,
  $i \in \dseto{m}$,
  are draws from independent random variables each with zero mean,
  \ie $\expe{\xi_i}= 0$,
  if the permanent and temporary impact functions are given by \eqref{eq:linear_permanent_impact}-\eqref{eq:linear_temporary_impact},
  and if the temporary impact parameter $\epsilon_{i}$ is constant over all trading periods,
  \ie $\forall i \in \dseto{m}: \epsilon_{i} = \epsilon$,
  then there exists a strategy $\bar{\ystratb}$ and a redistribution matrix $\bar{\betab} \mdef \pa{ \beta_{k,i} } \in \R^{m-1 \times m}$ such that the trading volumes given by \Cref{eq:relation_volume_proportion} are equal to $\rquantb^{rec}$.
\end{lemma}
\begin{proof}
  Let define $\bar{\ystratb}$ as $ \ystratb^{\star,t_0} > 0 $,
  the optimal solution of Optimisation Problem~\eqref{eq:mean_variance_optimisation_problem} at time~$t_0$,
  and
  $\bar{\betab} = \pa{ \bar{\beta}_{k,i} } \in \R^{m-1 \times m} $ 
  as the redistribution matrix implied by $\ystratb^{\star,t_0}$,
  \ie the matrix whose components are defined as follows
  \begin{equation}
    \label{eq:implied_redist_matrix}
    \bar{\beta}_{k,i} \pa{ \ystratb^{\star,t_0} }
    \mdef
    \begin{cases}
      0 & \text{ if } i \leq k, \\
      \frac{\ystrat^{\star,t_0}_{i}}{\sum_{r=k+1}^{m}{\ystrat^{\star,t_0}_{r}}}
	& \text{ otherwise. }
    \end{cases}
  \end{equation}
  Then,
  $\bar{\ystratb}$ and $\bar{\betab}$ are such that the trading volumes given by \Cref{eq:relation_volume_proportion} are equal to $\rquantb^{rec}$.
  Indeed,
  using the fact that $\sum_{i=1}^{m} \ystrat^{\star,t_{0}}_{i} = 1$,
  one has that the volumes $\rquantb^{rec}$ are given by

  \begin{align}
    \rquant_{1}^{rec} & = \ystrat^{\star,t_{0}}_{1} \rtarget_{0}, \\
    \rquant_{2}^{rec} &
    =
    \frac{\ystrat^{\star,t_{0}}_{2}}{\sum_{r=2}^{m} \ystrat^{\star,t_{0}}_{r}} \pa{\rtarget_{0} + \rupdate_{1} - \ystrat^{\star,t_{0}}_{1} \rtarget_{0}}
    =
    \frac{\ystrat^{\star,t_{0}}_{2}}{1 - \ystrat^{\star,t_{0}}_{1}} \pa{ \pa{ 1 - \ystrat^{\star,t_{0}}_{1} } \rtarget_{0} + \rupdate_{1} } \nonumber \\
  &
    =
    \ystrat^{\star,t_{0}}_{2} \rtarget_{0} + 
    \frac{\ystrat^{\star,t_{0}}_{2}}{1 - \ystrat^{\star,t_{0}}_{1}} \rupdate_{1} \\
    \rquant_{3}^{rec} &
    =
    \frac{\ystrat^{\star,t_{0}}_{3}}{\sum_{r=3}^{m} \ystrat^{\star,t_{0}}_{r}} \pa{
  \rtarget_{0} + \rupdate_{1} + \rupdate_{2}
  - \ystrat^{\star,t_{0}}_{1} \rtarget_{0}
  - \ystrat^{\star,t_{0}}_{2} \rtarget_{0}
  - \frac{\ystrat^{\star,t_{0}}_{2}}{1 - \ystrat^{\star,t_{0}}_{1}} \rupdate_{1}
    } \nonumber \\
  &
    = 
    \ystrat^{\star,t_{0}}_{3} \rtarget_{0}
    + \frac{\ystrat^{\star,t_{0}}_{3}}{1 - \ystrat^{\star,t_{0}}_{1}} \rupdate_{1}
    + \frac{\ystrat^{\star,t_{0}}_{3}}{1 - \ystrat^{\star,t_{0}}_{1} - \ystrat^{\star,t_{0}}_{2}} \rupdate_{2} \\
  & \vdots \nonumber \\
    \rquant_{i}^{rec} &
    = 
    \ystrat^{\star,t_{0}}_{i} \rtarget_{0}
    + \sum_{k=1}^{i-1} 
    \pa{
  \frac{\ystrat^{\star,t_{0}}_{i}}{ 1 - \sum_{r=1}^{k} \ystrat^{\star,t_{0}}_{r}}
    } \rupdate_{k}. \label{eq:rec_volume_i}
  \end{align}
  Using the definitions of $\bar{\ystratb}$ and $\bar{\betab}$,
  one obtains
  \begin{equation}
    \rquant_{i}^{rec}
    =
    \bar{\ystrat}_{i} \rtarget_{0}
    + \sum_{k=1}^{i-1} 
    \pa{
      \frac{ \pa{1 - \sum_{r=1}^{k} \bar{\ystrat}_{r} } \bar{\ystrat}_{i} + \bar{\ystrat}_{i} \pa{\sum_{r=1}^{k} \bar{\ystrat}_{r} } }{ 1 - \sum_{r=1}^{k} \bar{\ystrat}_{r}}
    } \rupdate_{k}
    = 
    \bar{\ystrat}_{i} \rtarget_{0}
    + \sum_{k=1}^{i-1} 
    \pa{
      \bar{\ystrat}_{i}
      +
    \bar{\beta}_{k,i} \sum_{r=1}^{k} \bar{\ystrat}_{r} } \rupdate_{k},
  \end{equation}
  which is exactly the expression of the traded volumes in \eqref{eq:mean_CVaR_optimisation_problem} for the strategy $\bar{\ystratb}$ and the redistribution matrix~$\bar{\betab}$ (see \Cref{eq:relation_volume_proportion}).
\end{proof}%

\noindent
\Cref{lemma:mean_CVaR_reproducibility_mean_variance_with_recourse} implies that,
if we were to optimise~\eqref{eq:mean_CVaR_optimisation_problem} on the set of strategies $\mset{Y}$ and the set of redistribution matrices $\mset{B} \mdef \big\{ \betab \in \R^{m-1 \times m} \, \big| \, \beta_{k,i} = 0  \text{ if } i\leq k \text{ and } \forall k \in \dseto{m-1}: \sum^{m}_{i=k+1} \beta_{k,i} = 1 \big\}$,
then there exists a strategy $\pa{\ystratb,\betab} \in \mset{Y} \times \mset{B}$ such that the volumes traded by following this strategy are the same as the ones of a trader recomputing the optimal strategy in the mean-variance framework whenever an update on the volume target is released.
Hence,
a wisely chosen redistribution matrix $\betab$ can mimic the action of taking recourse.
A sensible choice for the redistribution matrix~$\betab$ would for instance be the matrix implied by the optimal risk-neutral strategy $\ystratb^{\star, t_{0}, \lambda_{\Var} = 0}$ in the mean-variance framework
(Optimisation Problem~\eqref{eq:mean_variance_optimisation_problem}) at time $t_{0}$,
\ie the matrix defined as in \eqref{eq:implied_redist_matrix} but with $\ystratb^{\star, t_{0}, \lambda_{\Var} = 0}$.

\paragraph{Exponential utility and mean-QV.}%
\label{par:exponential_utility_and_mean_qv}

To the best of our knowledge,
considering the uncertainty of the volume target has so far been overlooked in the literature.
Some articles deal with a conceptually similar problem to ours but rely on assumptions that are discarded in our model,
allowing for greater generality as explained hereinafter.
In \citet{Cheng2017,Cheng2019},
the authors incorporate in the optimal execution problem the uncertainty related to order fills,
which is the risk for an order to be under- or over-executed.
\citet{Bulthuis2017} extend this model by including,
additionally to the market orders, limit orders with uncertain fill rates.
The authors also add penalties to tailor the orders sign and orders magnitude.
For comparison,
we consider the model of \citet{Cheng2017} as we make similar assumptions on the market;
the model is here formulated as a liquidation problem,
\ie from a seller's perspective rather than the buyer's perspective adopted earlier. 
Their model is as follows.
Let $x_{t}$ denote the number of shares that the trader holds at a time $t \in \cset{0}{T}$ during the execution process;
 $x_{0}$ denotes the initial amount of shares.
Similarly to \citet{Almgren2001},
the fair price $\rprice_{t}$ is driven by the SDE
\begin{equation}
  \label{eq:cheng_sde}
\dif \rprice_{t} = \gamma \dif x_{t} + \mu \dif t + \sigma \dif W_{t}
\Leftrightarrow 
\rprice_{t} = \rprice_{0} - \gamma x_{0} + \gamma x_{t} + \mu t + \sigma W_{t}
\end{equation}
and the transacted price $\tilde{\rprice}_{t}$ is given by
\begin{equation}
  \label{eq:cheng_trans_price}
\tilde{\rprice}_{t} = \rprice_{t} - \eta v_{t},
\end{equation}
where $W_{t}$ is a Brownian motion,
$v_{t}$~denotes the rate at which the trader scheduled to trade at the instant~$t$,
$\mu$~parametrises the price drift,
and~$\gamma x_{t}$, 
$\gamma \geq 0$,
and~$-\eta v_{t}$,
$\eta > 0$,
respectively refer to the permanent and temporary impact of the trades.
To take into account the uncertainty regarding the order fills,
\citet{Cheng2017} add to the dynamics of the position $x_{t}$ a noise driven by another Brownian motion $Z_{t}$ (potentially correlated with $W_{t}$),
\ie
\begin{equation}
  \label{eq:cheng_dynamic_position}
  \dif x_{t} = - v_{t} \dif t + m \spa{v_{t}} \dif Z_{t},
\end{equation}
where the diffusion $m\spa{v_{t}}$ characterises the magnitude of the uncertainty of order fills.
At each time $t \in \cset{0}{T}$,
the profit and loss (\pnl) $\Pi_{t}\pa{x}$ generated by the strategy $x$ till time $t$ is
\begin{equation}
  \label{eq:cheng_pnl}
  \Pi_{t}\pa{x}
   =
  x_{t} \pa{\rprice_{t} - \rprice_{0}} - \int_{0}^{t} \pa{\tilde{\rprice}_{u} - \rprice_{0}} \dif x_{u}.
\end{equation}
The first term of \eqref{eq:cheng_pnl} represents the change in fair value of the positions that still have to be transacted (it is beneficial for the trader if $\rprice_{t} > \rprice_{0}$),
whereas the second term is the cost incurred due to the price risk and price impact of the shares traded up to time $t$.
Using \cref{eq:cheng_sde,eq:cheng_trans_price},
the \pnl can be reformulated as
\begin{equation}
  \label{eq:cheng_pnl_final}
  \Pi_{t}\pa{x}
   =
  x_{t} \pa{\gamma \pa{x_{t} - x_{0}} + \mu t + \sigma W_{t}}
  -
  \int_{0}^{t} \pa{\gamma \pa{x_{u} - x_{0}} + \mu u + \sigma W_{u} - \eta v_{u}} \dif x_{u}.
\end{equation}
To discourage any residual positions at the terminal time $T$,
a penalty term $f: x_{T} \to - \beta x_{T}^2$,
$\beta \geq \eta > \frac{\gamma}{2}$,
is added to $\Pi_{T}\pa{x}$,
which gives the final \pnl $\widetilde{\Pi}_{T}\pa{x} \mdef \Pi_{T}\pa{x} + f\pa{x_{T}}$.
In the case where the magnitude of the uncertainty of order fills is constant ($m\spa{v_{t}} = m_{0}$)
and where there is no drift ($\mu = 0$),
\cite{Cheng2017} provide closed form expressions for the optimal trading rate $v_{t}^{\star}$ and liquidation trajectory $x_{t}^{\star}$ of a risk-averse trader assessing risk in two distinct ways.
On the one hand,
by maximising the expected utility of the final \pnl,
\ie $\sup_{v \in \mset{A}} \expe{U\pa{\widetilde{\Pi}_{T}\pa{x}}}$ with $U$ the exponential utility function:
$U\pa{x} = \frac{1 - e^{- \theta_{\EU} x}}{\theta_{\EU}}$.
On the other hand,
by maximising the expected \pnl penalised by its quadratic variation \citep{Forsyth2012},
\ie $\sup_{v \in \mset{A}} \expe{ \widetilde{\Pi}_{T}\pa{x} - \lambda_{\QV} \quadvari{\Pi_{T}\pa{x}}}$.
In both cases,
$\mset{A}$ denotes the set of admissible controls.
The risk-aversion parameter are respectively given by $\theta_{\EU} > 0$ and $\lambda_{\QV} > 0$;
the larger $\theta_{\EU}$ (\resp $\lambda_{\QV}$),
the more risk-averse the trader.

Based on the optimal trading rate $v_{t}^{\star}$ given by \citet{Cheng2017},
it is straightforward to obtain the optimal trading rate of the execution problem formulated from a buyer perspective desiring to acquire an uncertain amount of shares $\rtarget_{T}$.
Indeed,
the fluctuations in the number of positions of the trader due to the uncertain order fills can be seen as fluctuations in the uncertain volume target.
Hence,
if the dynamics of the volume target forecast $\rtarget_{t} \mdef \cexpe{\rtarget_{T}}{\mset{F}_{t}}$ is given by $\dif \rtarget_{t} = m_{0} \dif X_{t}$,
where $X_{t}$ is a Brownian motion and $\pa{\mset{F}_{t}}_{t \in \cset{0}{T}}$ is the filtration of sigma algebras that represent the information available at time~$t$,
if~$w_{t}$ denotes the trading rate at the instant $t$,
and if~$z_{t}$ denotes the number of shares left to trade in order to match with the current forecast~$\rtarget_{t}$ on the volume target,
then the dynamics of~$z_{t}$ is given by
\begin{equation}
  \label{eq:dynamic_position_left}
  \dif z_{t} = - w_{t} \dif t + m_{0} \dif X_{t},
\end{equation}
where the initial amount of shares to trade $z_{0}$ is equal to $\rtarget_{0} \mdef \cexpe{\rtarget_{T}}{\mset{F}_{0}}$.
Note that, in contrast to~$v_{t}$,
a positive value for~$w_{t}$ corresponds to the acquisition of shares.
The impact of the trades on the price dynamics and the transacted price is then in the opposite direction compared to the model of \citet{Cheng2017}.
The price drift and price risk are also taken in the opposite direction compared to \eqref{eq:cheng_sde} given the rationale that a positive price shift in a liquidation program has a similar beneficial effect as a negative price shift in an acquisition program.
The fair price $\rprice_{t}$ dynamics,
the transacted price $\tilde{\rprice}_{t}$,
and the \pnl $\Pi_{t}\pa{z}$ are then given by

\begin{align}
  \dif \rprice_{t}    & = - \gamma \dif z_{t} - \mu \dif t - \sigma \dif W_{t}
  \Leftrightarrow 
  \rprice_{t} = \rprice_{0} + \gamma \pa{z_{0} - z_{t}} - \mu t - \sigma W_{t},
  \label{eq:volume_sde}
  \\
  \tilde{\rprice}_{t} & = \rprice_{t} + \eta w_{t},
  \label{eq:volume_trans_price}
  \\
  \Pi_{t}\pa{z}
		& =
		z_{t} \pa{\rprice_{0} - \rprice_{t}} - \int_{0}^{t} \pa{\rprice_{0} - \tilde{\rprice}_{u}} \dif z_{u}. \label{eq:volume_pnl}
\end{align}
Similarly to \eqref{eq:cheng_pnl},
the first term of \eqref{eq:volume_pnl} represents the change in fair value of the positions that still have to be acquired (it is beneficial for the trader if $\rprice_{t} < \rprice_{0}$),
whereas the second term is the cost incurred due to the price risk and price impact of the shares traded up to time $t$.
Using \cref{eq:volume_sde,eq:volume_trans_price},
the \pnl can be reformulated as
\begin{equation}
  \label{eq:volume_pnl_final}
  \Pi_{t}\pa{z}
   =
  z_{t} \pa{\gamma \pa{z_{t} - z_{0}} + \mu t + \sigma W_{t}}
  -
  \int_{0}^{t} \pa{\gamma \pa{z_{u} - z_{0}} + \mu u + \sigma W_{u} - \eta w_{u}} \dif z_{u}.
\end{equation}
We observe that \cref{eq:cheng_pnl_final,eq:volume_pnl_final} are similar.
Therefore,
if a risk-averse trader either maximises the expected utility of the final \pnl,
\ie $\widetilde{\Pi}_{T}\pa{z} = \Pi_{T}\pa{z} + f\pa{z_{T}}$,
either maximises the expected \pnl penalised by its quadratic variation,
then the optimal trading rate $w_{t}^{\star}$ in an acquisition program subjected to an uncertain volume target,
will be the same as the optimal trading rate $v_{t}^{\star}$ in a liquidation program subjected to uncertain order fills.
If additionally,
we assume the absence of price drift and that any remaining positions is prohibited,
\ie $\mu = 0$ and $\beta \to \infty$,
then the continuous execution problem formulated here above tackles essentially the same problem as Model~\eqref{eq:mean_CVaR_optimisation_problem} in the absence of fixed cost,
\ie $\forall i \in \dseto{m}: \epsilon_{i} = 0$,
when the price shifts~$\rpriceshift_{i}$ and forecast updates~$\rupdate_{i}$,
$i \in \dseto{m}$,
follow respectively a zero mean normal distribution of variance $\tau_{i} \sigma^{2}$ and $\tau_{i} m_{0}^{2}$,
\ie
$\rpriceshift_{i} \sim \normald{0}{\tau_{i} \sigma^{2}}$
and
$\rupdate_{i} \sim \normald{0}{\tau_{i} m_{0}^{2}}$.

To conclude,
there are multiple advantages of our approach.
Firstly,
we showed in \Cref{lemma:mean_CVaR_reproducibility_mean_variance_with_recourse} that the traded volumes resulting of the strategy consisting of recomputing the optimal strategy in the mean-variance framework whenever a volume target update becomes available can be reproduced in our model with an appropriate updating rule $\betab$.
Secondly,
our model is significantly more versatile compared to the model of \citet{Cheng2017}.
Indeed,
to provide closed form expressions of the optimal trading rate,
\citet{Cheng2017} rely on the assumptions that both the price dynamics and the uncertain order fills are modelled with Brownian motions,
as well as on the assumption that the price volatility ($\sigma$),
the volatility of the updates on the volume target ($m_{0}$),
and the price impact parameters ($\gamma$ and $\eta$) are constant over the entire execution period.
All these considerations are not needed in our model,
which allows us to consider a wider range of situations,
such as,
for instance,
the case where the liquidity profile is not constant over the course of the execution period or the situation where the forecast updates on the volume target are correlated.
In these situations,
finding closed form expressions of the optimal trading rate for the model proposed by \citet{Cheng2017} would be significantly more challenging.
Alternatively,
the optimal rate could be obtained using a dynamic programming approach;
however,
this would result in a significantly higher computational cost compared to our approach. 
Finally,
as we will show in \Cref{sec:numerical_results},
we will observe that,
despite the fact that the $\CVaR_{\alpha}$ risk measure is not time-consistent which may lead to sub-optimal strategies \citep{Rudloff2014},
the strategies obtained with our model have competitive performance compared to the optimal solution provided by the dynamical programming approach suggested in \cite{Cheng2017}. 
This suggests that integrating the uncertainty related to the volume target in the estimate of the trading cost of a strategy (see \Cref{eq:mean_CVaR_trading_cost})
and taking partial recourse via a predetermined rule $\betab$
lead to intuitive and competitive trading strategies while avoiding the computational cost of more convoluted approaches.

%% file: sections/3-numerical_results.tex
\section{Numerical results}%
\label{sec:numerical_results}

In this section,
we analyse the influence of incorporating the uncertainty related to the volume target into the trade execution problem.
We provide some numerical evidence that,
when the uncertainty related to the volume target is considered,
a trader should adapt their execution program by delaying their trades.
To illustrate the results,
we apply our model in two distinct test cases.
In both cases,
it is assumed that the market has a single stock \comadd{with initial price $\rprice_{0} = 50$},
and with median daily trading volume~$V$ of 5 million shares:
\begin{enumerate}
  \item[\textcolor{myOxfordBlue}{(a)}] \textbf{Constant bid-ask spread.} 
    We assume that the market liquidity remains constant over all the trading periods;
    \comrefo{we assume a bid-ask spread $b_{i}$ equal to 0.25\% of the initial asset price $\rprice_{0}$ for each trading period $\tau_{i}$,
    \ie $\forall i \in \dseto{m}: b_{i} = 0.125$.}
  \item[\textcolor{myOxfordBlue}{(b)}]
    \comrefo{
      \textbf{U-shaped intra-day bid-ask spread profile.}
      Previous papers,
      \eg \citet{Chan1995},
      suggest that the traded volumes and bid-ask spreads of NYSE stocks often follow a U-shaped pattern over the trading day;
      the trading activity being concentrated at the market opening and closure.
      Consequently,
      the price-impact slopes,
      \ie $\eta_{i}$,
      are larger in the early and last periods than in the middle periods \citep{Huberman2005}.
      To capture such behaviour,
      let assume the following profile for the bid-ask spread over the 5 trading periods:
      $\bb \mdef \mvec{b_{1}}{b_{5}}^{\T} = \pac{0.15, 0.125, 0.125, 0.2,  0.25}^{\T}$,
      which correspond to 0.3\%, 0.25\%, 0.25\%, 0.4\% and 0.5\% of the initial asset price $\rprice_{0}$.
    }
\end{enumerate}

For both cases,
\comreft{
  we assume that the volatility $\sigma$ of the asset is constant during the entire execution period and is equal to $0.95$,
  and that the price shift $\rpriceshift_{i}$,
  $i \in \dset{1}{m}$,
  follows a zero mean Gaussian distribution of variance $\tau_{i}\sigma^{2}$,
  \ie $\rpriceshift_{i} \sim \normald{0}{\tau_{i}\sigma^{2}}$.
}%
For the temporary impact,
the fixed cost factor $\epsilon_{i}$ is assumed to amount to one half of the bid-ask spread for all trading period $\tau_{i}$,
\ie $\forall i \in \dset{1}{m}: \epsilon_{i} = 0.5b_{i}$.
Additionally,
\citet{Almgren2001} suggest that a trader incurs a price impact equal to one bid-ask spread for every percent of the daily volume,
\ie $\forall i \in \dset{1}{m}: \eta_{i} = b_{i}/(0.01\cdot V)$,
which we adopt in our paper as well.
As for the permanent price impact,
\citet{Almgren2001} assume that the price effects become significant when 10\% of the daily volume is traded.
In this application,
we suggest that the effect is significant if it corresponds to 5 times the bid-ask spread of the trading period,
whereas \citet{Almgren2001} consider
a price shift equal to one bid-ask spread as significant.
This increase of scale allows us to illustrate more clearly the contribution of volume uncertainty in the risk adopted by the trader,
but it is also not completely unrealistic for illiquid assets.
We therefore set \comadd{$\gamma_{i} = \pa{5 b_{i}}/\pa{0.1 V}$ for all $i \in \dseto{m}$}.
For smaller values of~$\gamma_{i}$,
the impact of volume uncertainty is slightly smaller but,
due to the cumulative effect of sequential trades, 
even small savings in the trading cost make a large difference over time.

Finally,
we assume \comadd{
  for both cases that the initial volume target $\rtarget_{0}$ amounts to 0.5 million shares,
  and that for each trading period~$\tau_{i}$ except the last one,
  the forecast update follows a zero mean Gaussian distribution of variance $\tau_{i} \nu^{2}$ with $\nu$ being equal to 10\% of the initial volume target ($\nu \mdef 0.1 \rtarget_{0}$),
  \ie $\forall i \in \dset{1}{m-1}: \rupdate_{i} \sim \normald{0}{\tau_{i} \nu^2}$.
}
As suggested in~\Cref{sub:frameworks_comparison_theory},
we assume that the residual volumes to trade due to the forecast updates are redistributed over the future trading periods based on the matrix~$\betab$ implied by the risk-neutral optimal trading strategy~$\ystratb^{\star, t_{0}, \lambda_{\Var} = 0}$ in the mean-variance framework (Optimisation Problem~\eqref{eq:mean_variance_optimisation_problem}) computed at time $t_{0}$,
\ie
$ \betab = \pa{ \beta_{k,i} \pa{ \ystratb^{\star, t_{0}, \lambda_{\Var} = 0} } } \in \R^{m-1 \times m} $ with
\begin{equation}
  \label{eq:red_mat_num_res}
  \beta_{k,i} \pa{ \ystratb^{\star, t_{0}, \lambda_{\Var} = 0} } = 
  \begin{cases}
    0 & \text{ if } i \leq k, \\
    \frac{\ystrat^{\star, t_{0}, \lambda_{\Var} = 0}_{i}}{\sum_{r=k+1}^{m}{\ystrat^{\star, t_{0}, \lambda_{\Var} = 0}_{r}}}
      & \text{ otherwise. }
  \end{cases}
\end{equation}
In practice,
the exact distributions of the price moves and the forecast updates are unknown and are estimated from historical data (\eg see \Cref{sub:application_power}).
In the following,
we denote by~$\hat{\cdot}$ any trader's estimation of the ground truth parameter.
For example,
$\tau_{i}\hat{\nu_{i}}^{2}$ is the trader estimated value of~$\tau_{i}\nu_{i}^{2}$,
the variance of the forecast update released after trading period~$\tau_{i}$.
\Cref{tab:market_parameters} summarises the market details of both cases.

\begin{table}[h]
  \begin{center}
    {\footnotesize
      \caption{Market parameters}
      \label{tab:market_parameters}
      \begin{tabular}{|rl|}
	\hline
	& \\[-8pt]
	Acquisition time: & $T = 5$ [day]\\
	Number of trading periods: & $m = 5$\\
	Trading period length: & $\forall i \in \dset{1}{m}: \tau_{i} = T/m = 1$ [day]\\
	Initial asset price:  &$\rprice_{0} = 50$ [\$/share] \\
	\comreft{Asset's volatility}: & $\sigma$ = 0.95 [\si{(\$ \per share) \per day^{1/2}}] \\
	\comreft{Price shifts distribution}: & $\forall i \in \dset{1}{m}: \rpriceshift_{i} \sim \normald{0}{\tau_{i}\sigma^{2}}$\\
	Daily volume: & $V = 5 \cdot 10^6$ [share]\\
	\comadd{Initial volume target forecast}: & $\rtarget_{0} = 0.5 \cdot 10^6$ [\si{share}]\\
	Volume forecast updates distribution: & $\forall i \in \dset{1}{m-1}: \rupdate_{i} \sim \normald{0}{\tau_{i} \nu^2}$ with $\nu = 0.1 \rtarget_{0}$\\
	Bid-ask spread: & \textcolor{myOxfordBlue}{Case (a)}: $\forall i \in \dset{1}{m}: b_{i} = \comrefo{0.125}$ [\$/share]\\
			& \textcolor{myOxfordBlue}{Case (b)}: \comrefo{$\bb = \pac{0.15, 0.125, 0.125, 0.2,  0.25} $} [\$/share]\\
	Fixed cost: & $\forall i \in \dset{1}{m}: \epsilon_{i} = 0.5b_{i}$ [\$/share]\\
	Impact at 1\% of market: & $\forall i \in \dset{1}{m}: \eta_{i} = b_{i}/(0.01\cdot V)$ [\si{(\$ \per share) \per (share \per day)}]\\
	Permanent impact parameter:& \comadd{$\forall i \in \dset{1}{m}: \gamma_{i} = (5 b_{i})/(0.1V)$} [\si{\$ \per share^2}]\\
	$\CVaR_\alpha$ parameter:& $\alpha = \comrefo{0.1}$ \\[4pt]
	\hline
      \end{tabular}
    }%
  \end{center}
\end{table}

\input{\pathSectionsOptimalTrading/3.1-optimal_strategies.tex}
\input{\pathSectionsOptimalTrading/3.2-model_performance.tex}
\input{\pathSectionsOptimalTrading/3.3-framework_comparison.tex}
\input{\pathSectionsOptimalTrading/3.4-power-forecast.tex}

%% file: sections/3.1-optimal_strategies.tex
\subsection{Optimal strategies and impact of the liquidity profile}%
\label{sub:optimal_strategies}

\begin{figure}
  \centering
  \begin{adjustbox}{minipage=\linewidth,scale=1}
    \subfloat[
    Case (a):
    optimal trading strategies under PU only,
    \ie~$\forall i \in \dseto{m-1}: \hat{\nu}_i = 0$.
    ]{\label{fig:price_uncertainty}\includegraphics[width=0.48\linewidth]{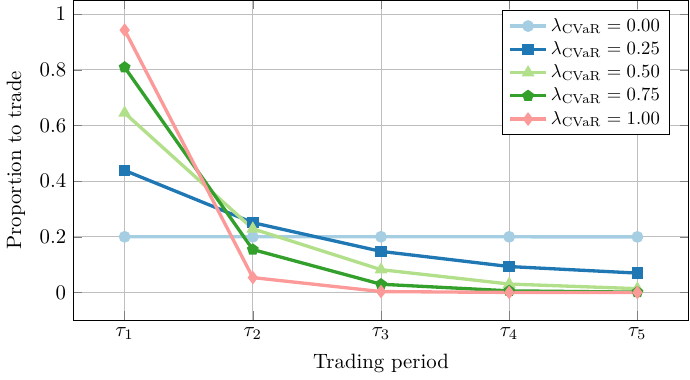}}\hspace*{\fill}
    \subfloat[
    Case (b):
    optimal trading strategies under PU only,
    \ie~$\forall i \in \dseto{m-1}: \hat{\nu}_i = 0$.
    ]{\label{fig:price_uncertainty_different_liquidity}\includegraphics[width=0.48\linewidth]{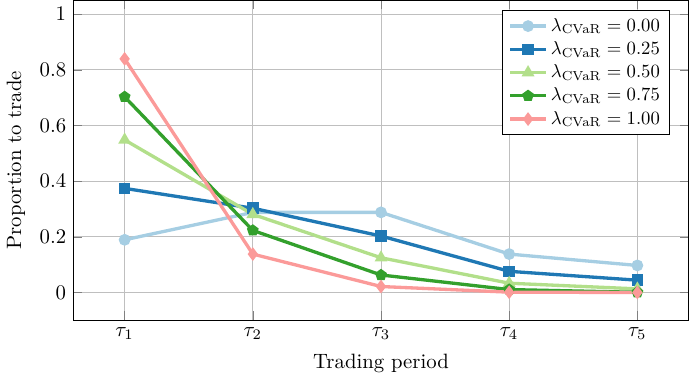}}\hspace*{\fill}\\
    \subfloat[
    Case (a):
    optimal trading strategies under PVU,
    \ie~$\forall i \in \dseto{m-1}: \hat{\nu}_i = \nu_i$.
    ]{\label{fig:price_volume_uncertainty}\includegraphics[width=0.48\linewidth]{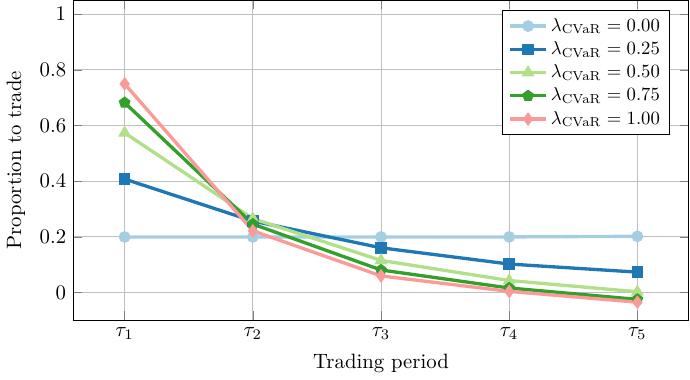}}\hspace*{\fill}
    \subfloat[
    Case (b):
    optimal trading strategies under PVU,
    \ie~$\forall i \in \dseto{m-1}: \hat{\nu}_i = \nu_i$.
    ]{\label{fig:price_volume_uncertainty_different_liquidity}\includegraphics[width=0.48\linewidth]{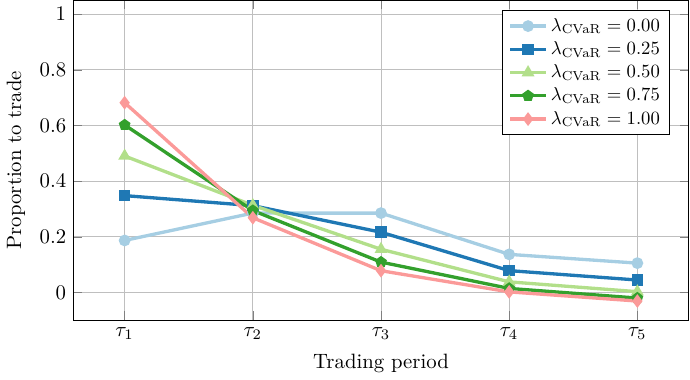}}\hspace*{\fill}\\
    \subfloat[
    Case (a):
    difference between the strategies represented in \Cref{fig:price_volume_uncertainty} with the ones represented in \Cref{fig:price_uncertainty}.
    ]{\label{fig:price_volume_uncertainty_mean_CVaR_impact}\includegraphics[width=0.48\linewidth]{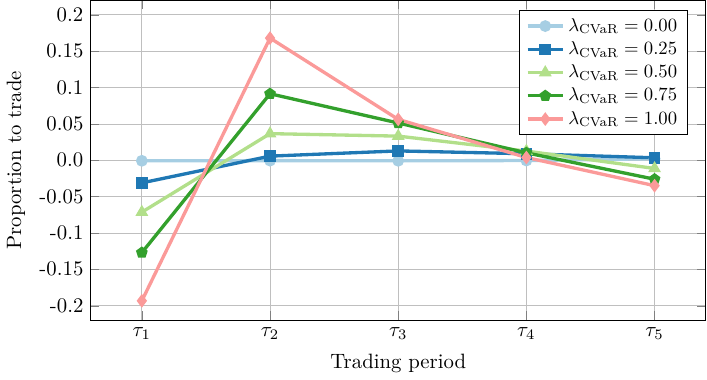}}\hspace*{\fill}
    \subfloat[
    Case (b):
    difference between the strategies represented in \Cref{fig:price_volume_uncertainty_different_liquidity} with the ones represented in \Cref{fig:price_uncertainty_different_liquidity}.
    ]{\label{fig:price_volume_uncertainty_mean_CVaR_impact_different_liquidity}\includegraphics[width=0.48\linewidth]{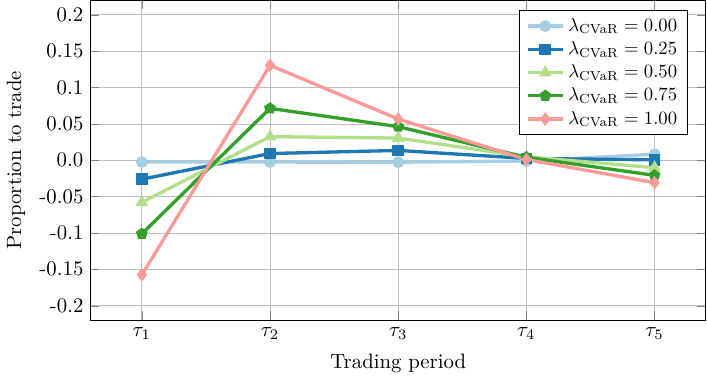}}\hspace*{\fill}
  \end{adjustbox}
  \caption{
    Optimal trading strategies under different assumptions for both test cases.
  }
  \label{fig:mean_CVaR_price_and_volume_uncertainty}
\end{figure}

\comadd{
  Given the market parameters of \Cref{tab:market_parameters},
  the condition $\Mb \succ 0$ of \Cref{theorem:mean_CVaR_convexity_model} is verified,
  which means that for any risk-aversion parameter $\lambda_{\CVaR} \in \pac{0,1}$,
  the optimal trading strategy in the Mean-CVaR framework is uniquely defined.
}%
\Cref{fig:price_uncertainty,fig:price_uncertainty_different_liquidity} depict,
for both liquidity profiles,
the optimal strategies of Model~\eqref{eq:mean_CVaR_optimisation_problem} when only price uncertainty (PU) is considered,
whereas
\Cref{fig:price_volume_uncertainty,fig:price_volume_uncertainty_different_liquidity} depict,
the optimal trading strategies when both sources of uncertainty are taken into account,
\ie price and volume uncertainty (PVU).
The difference between the optimal strategies in these two situations is represented for both cases respectively in \Cref{fig:price_volume_uncertainty_mean_CVaR_impact,fig:price_volume_uncertainty_mean_CVaR_impact_different_liquidity}.
This difference illustrates how the optimal strategies under price uncertainty only should be adjusted when a trader additionally takes into account the uncertainty related to the volume target.

The impact of considering the uncertainty related to the volume to trade on the optimal strategies is consistent with intuition:
if the volume to trade is uncertain,
waiting to receive a more precise estimate of this volume is advantageous in terms of risk reduction.
Indeed,
we observe that a trader should mitigate their traded volume of the first trading period and spread it over the next periods.
Moreover,
the more risk-averse the trader,
the bigger the impact on their optimal strategy.

These figures also illustrate that the impact of considering volume uncertainty on the optimal execution strategy is non-monotonic due to two different subsets of extreme events:
those in which volume forecasts are increased in late trading periods,
and those where volume forecasts are downsized.
The first situation requires carrying out enough early trades to avoid having to upsize the traded volume in the latest stages to a level that is strongly affected by liquidity constraints.
The second situation makes it necessary to avoid over-trading in the early trading periods and having to trade out of these positions later.
The two effects tend to counterbalance each other,
with the result of reallocating some of the early trades to the middle section of the trading period.
This effect is picked up due to the incorporation of recourse estimates and the use of the $\CVaR_{\alpha}$ measure that focuses on extreme events.

\begin{figure}
  \centering
  \begin{adjustbox}{minipage=\linewidth,scale=1.0}
    \subfloat{\label{fig:liquidity_profile_shape_1} \includegraphics[width=0.19\linewidth]{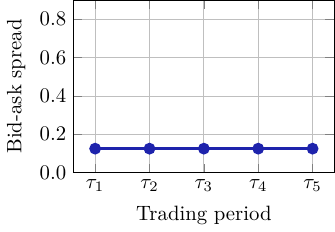}}\hspace*{\fill}
    \subfloat{\label{fig:liquidity_profile_shape_2} \includegraphics[width=0.19\linewidth]{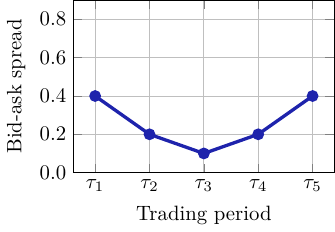}}\hspace*{\fill}
    \subfloat{\label{fig:liquidity_profile_shape_3} \includegraphics[width=0.19\linewidth]{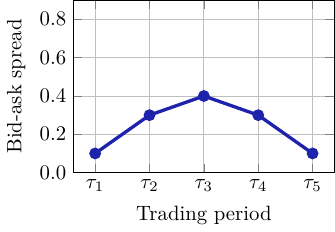}}\hspace*{\fill}
    \subfloat{\label{fig:liquidity_profile_shape_4} \includegraphics[width=0.19\linewidth]{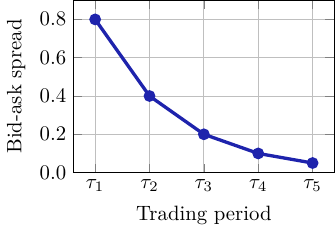}}\hspace*{\fill}
    \subfloat{\label{fig:liquidity_profile_shape_5} \includegraphics[width=0.19\linewidth]{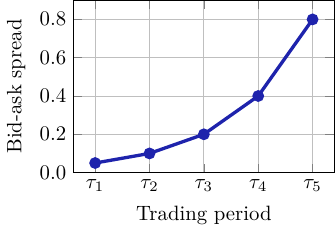}}\\
    \subfloat{\label{fig:liquidity_profile_PVU_1}   \includegraphics[width=0.19\linewidth]{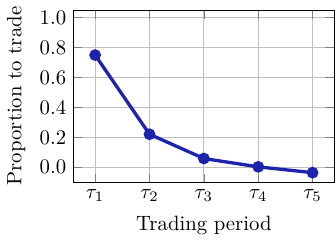}}\hspace*{\fill}
    \subfloat{\label{fig:liquidity_profile_PVU_2}   \includegraphics[width=0.19\linewidth]{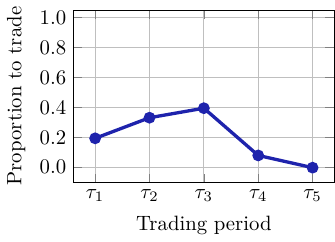}}\hspace*{\fill}
    \subfloat{\label{fig:liquidity_profile_PVU_3}   \includegraphics[width=0.19\linewidth]{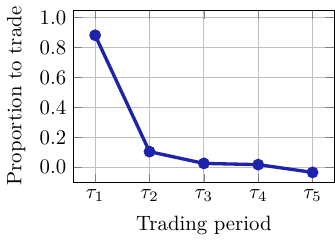}}\hspace*{\fill}
    \subfloat{\label{fig:liquidity_profile_PVU_4}   \includegraphics[width=0.19\linewidth]{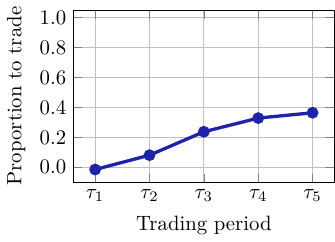}}\hspace*{\fill}
    \subfloat{\label{fig:liquidity_profile_PVU_5}   \includegraphics[width=0.19\linewidth]{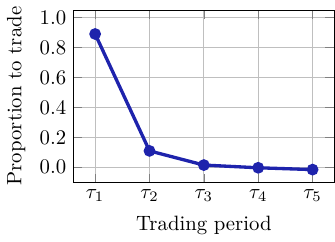}}\hspace*{\fill}\\
    \subfloat{\label{fig:liquidity_profile_diff_1}  \includegraphics[width=0.19\linewidth]{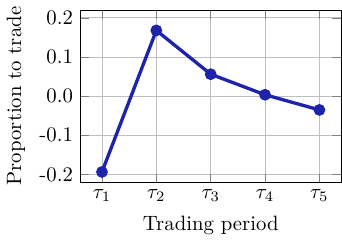}}\hspace*{\fill}
    \subfloat{\label{fig:liquidity_profile_diff_2}  \includegraphics[width=0.19\linewidth]{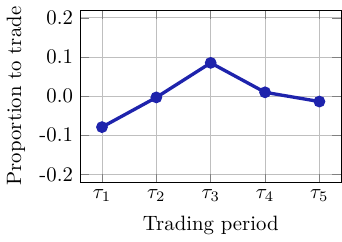}}\hspace*{\fill}
    \subfloat{\label{fig:liquidity_profile_diff_3}  \includegraphics[width=0.19\linewidth]{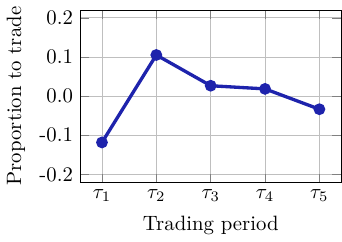}}\hspace*{\fill}
    \subfloat{\label{fig:liquidity_profile_diff_4}  \includegraphics[width=0.19\linewidth]{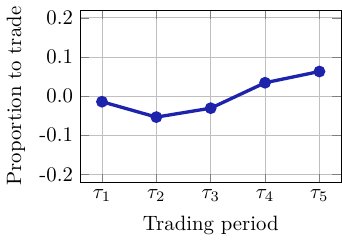}}\hspace*{\fill}
    \subfloat{\label{fig:liquidity_profile_diff_5}  \includegraphics[width=0.19\linewidth]{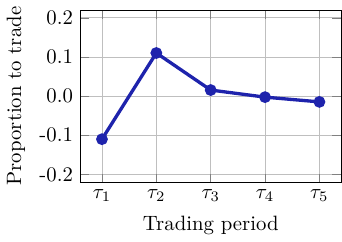}}\hspace*{\fill}
  \end{adjustbox}
  \caption{
    Analysis of the impact of the bid-ask spread profile on the strategy adjustments of an exclusively risk-focused trader ($\lambda_{\CVaR} =1$) when considering the uncertainty related to the volume target
    (row 1: bid-ask spread profile,
    row 2: optimal strategy in the mean-$\CVaR_{0.1}$ (PVU) framework,
    row 3: strategy adjustments between the strategies under PU and PVU).
  }
  \label{fig:analysis_liquidity_profile_impact}
\end{figure}

\comrefo{
  The liquidity profile has an impact both on the optimal strategies and on the adjustments to make to the strategies when considering the uncertainty related to the volume target additionally to the price uncertainty.
  Firstly,
  compared to the situation where liquidity remains constant,
  a risk-neutral trader should shift their trade from less to more liquid periods.
  This is not as simple for risk-averse traders, since the more risk-averse a trader is,
  the larger the impact of adversarial scenarios on their objective function,
  and consequently,
  the more the trader factors in the impact of market liquidity in conjunction with uncertainty.
  The decreasing influence of the market liquidity is observed in \Cref{fig:price_uncertainty_different_liquidity,fig:price_volume_uncertainty_different_liquidity} as the shape of the risk-neutral strategy gradually vanishes with increasing risk-aversion parameter, in favour of front loading to cope with price uncertainty.
  Secondly,
  the liquidity profile also impacts the adjustments to make to the strategies when introducing volume uncertainty into the model as illustrated in
  \Cref{fig:price_volume_uncertainty_mean_CVaR_impact,fig:price_volume_uncertainty_mean_CVaR_impact_different_liquidity,fig:analysis_liquidity_profile_impact};
  \Cref{fig:analysis_liquidity_profile_impact} depicts the adjustments to make for several other liquidity profiles as well as the optimal strategies under PVU.
  To cope with volume uncertainty,
  a trader has to consider the two different subsets of extreme events previously described.
  On the one hand,
  the less liquid the last trading periods,
  the more sensitive the trader to the risk of having to upsize the traded volume of the last trading periods,
  and therefore the smaller the planned proportions of the volume to trade over these periods.
  On the other hand,
  a risk-averse trader considers the risk of over-trading by mitigating their trading quantity of the first trading period and redistributing it over the next periods.
  The less liquid the market during these next trading periods,
  the greater is the risk of over-trading,
  and therefore the more beneficial in terms of risk reduction is the decrease of the trading quantity of the first trading period.
  These figures also show that the adjustment in the trading proportion of the first trading period is redistributed onto the next trading periods according to the liquidity profile:
  the more liquid a trading period relatively to another,
  the more it is favourable to increase its planned trade volume allocation rather than the one of this other period. 
}

%% file: sections/3.2-model_performance.tex
\subsection{Model performance}%
\label{sub:model_performance}

\begin{figure}[h]
  \centering
  \begin{adjustbox}{minipage=\linewidth,scale=1}
    \hspace*{\fill}
    \subfloat[$\lambda_{\CVaR} = 0.25$
    ]{\label{fig:impact_VU_risk_aversion_0_25}\includegraphics[width=0.46\linewidth]{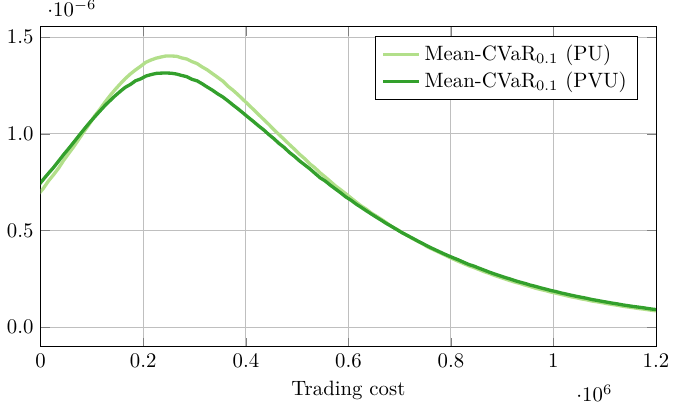}}\hspace*{\fill}
    \subfloat[$\lambda_{\CVaR} = 0.5$
    ]{\label{fig:impact_VU_risk_aversion_0_5}\includegraphics[width=0.46\linewidth]{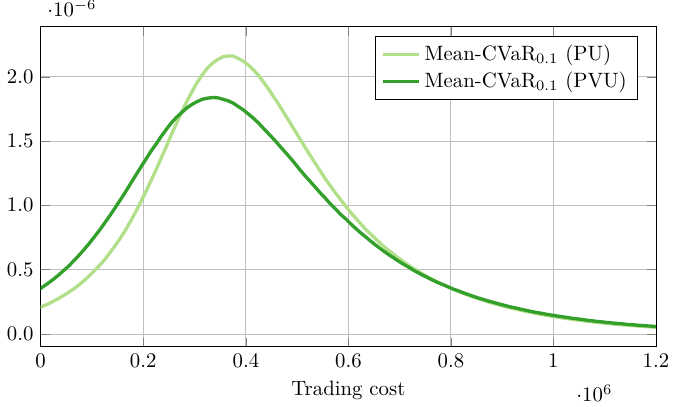}}
    \hspace*{\fill}
    \\
    \hspace*{\fill}
    \subfloat[$\lambda_{\CVaR} = 0.75$
    ]{\label{fig:impact_VU_risk_aversion_0_75}\includegraphics[width=0.46\linewidth]{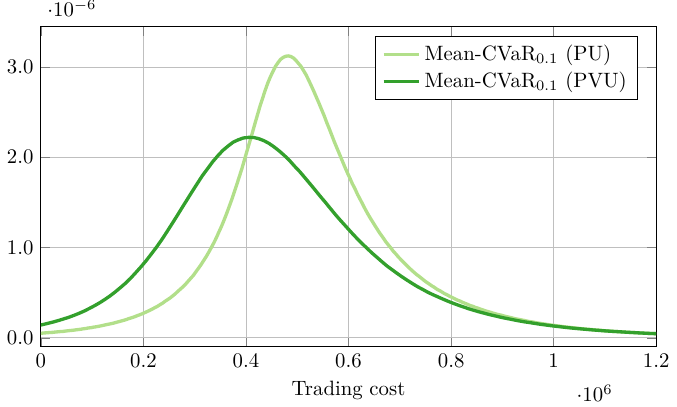}}\hspace*{\fill}
    \subfloat[$\lambda_{\CVaR} = 1.0$
    ]{\label{fig:impact_VU_risk_aversion_1_0}\includegraphics[width=0.46\linewidth]{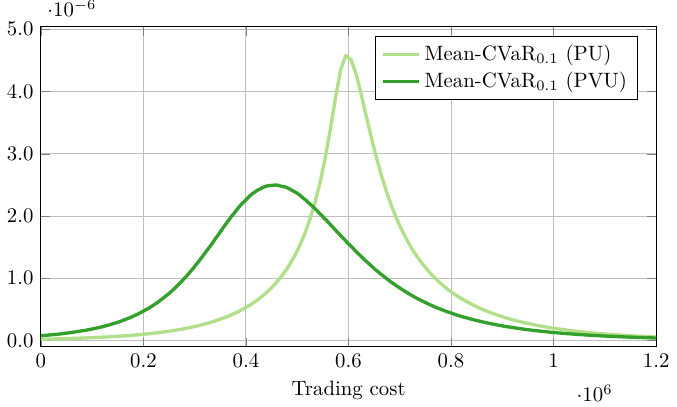}}
    \hspace*{\fill}
  \end{adjustbox}
  \caption{
    Comparison of the PDFs of the trading cost when the uncertainty related to the total volume to be traded is or is not taken into account in Model \eqref{eq:mean_CVaR_optimisation_problem}
    (a number of $10^8$ random paths were generated to simulate these empirical distributions).
  }
  \label{fig:impact_volume_uncertainty_on_trading_cost_densities}
\end{figure}

In this section,
we analyse the impact of considering volume uncertainty as part of the model on the strategy performance for Case~(a);
Case~(b) has similar observations.
We compare the performance of the strategies obtained when minimising the $\meanCVaR_{0.1}$ trade-off in two distinct cases.
The first case corresponds to a trader that wrongly assumes that the total volume to be traded is fixed and will not change along the course of the execution of their strategy,
\ie~$\forall i \in \dseto{m-1}: \hat{\nu}_i = 0$.
As a consequence,
the corresponding strategies are the ones obtained if the trader considers that uncertainty only arises from the price dynamics (see \Cref{fig:price_uncertainty} and \Cref{tab:mean_CVaR_without_volume_uncertainty}). 
The second case corresponds to the situation where the trader estimates correctly the variability of the volume target forecasts,
\ie~$\forall i \in \dseto{m-1}: \hat{\nu}_i = \nu_i$,
and takes it into account to define their trading strategy (see \Cref{fig:price_volume_uncertainty} and \Cref{tab:mean_CVaR_with_volume_uncertainty}).
\Cref{tab:mean_CVaR} reports for both cases the expectation, 
the $\alpha$-Conditional Value-at-Risk for $\alpha = 0.1$, 0.05 and 0.01,
and the variance of the trading cost incurred when following the strategies obtained by solving Model~\eqref{eq:mean_CVaR_optimisation_problem}.
\Cref{fig:impact_volume_uncertainty_on_trading_cost_densities} compares the probability density functions (PDFs) of the trading cost of both situations.
The more risk-averse a trader, 
the more pronounced the difference between the densities.
\Cref{tab:mean_CVaR} and \Cref{fig:impact_volume_uncertainty_on_trading_cost_densities} clearly illustrate the benefit of including the uncertainty related to the volume target in the $\CVaR_\alpha$ equivalent formulation of the return-risk trade-off model of \cite{Almgren2001}:
a trader assessing correctly the uncertainty of the volume target achieves better $\meanCVaR_{0.1}$ trade-offs.

\begin{table}[h]
  \caption{
    Performance of the strategies obtained by solving Optimisation Problem~\eqref{eq:mean_CVaR_optimisation_problem} for different risk-aversion parameter values $\lambda_{\CVaR}$.
  }
  \label{tab:mean_CVaR}
  \centering
  \subfloat[
  A trader considers that the only source of uncertainty is the price dynamics (strategies under PU only depicted in \Cref{fig:price_uncertainty}).
  ]{\label{tab:mean_CVaR_without_volume_uncertainty}
  \begin{adjustbox}{minipage=\linewidth,scale=0.8}
  \centering
  \begin{tabular}[t]{|c|ccccc||c|}
      \hline
      &&&&&& \\[-12pt]
      $\lambda_{\CVaR}$ [1] & Expectation [\$] & $\CVaR_{0.1}$ [\$] & $\CVaR_{0.05}$ [\$] & $\CVaR_{0.01}$ [\$] &Variance [$\$^2$] & $\varphi_{0.1}^{\lambda_{\CVaR}}$ [\$] \\[2pt]
      \hline
      0.00 & 2.973e5 & 1.351e6 & 1.586e6 & 2.093e6 & 3.085e11 & 2.973e5 \\
      0.25 & 3.412e5 & 1.062e6 & 1.242e6 & 1.641e6 & 1.334e11 & 5.216e5 \\
      0.50 & 4.303e5 & 9.445e5 & 1.088e6 & 1.419e6 & 6.393e10 & 6.874e5 \\
      0.75 & 5.268e5 & 9.320e5 & 1.055e6 & 1.349e6 & 3.930e10 & 8.307e5 \\
      1.00 & 6.273e5 & 9.833e5 & 1.096e6 & 1.368e6 & 3.143e10 & 9.833e5 \\
      \hline
    \end{tabular}
  \end{adjustbox}
  }\\
  \subfloat[
  A trader considers that the uncertainty arises from both the price dynamics and the total volume to be traded (strategies under PVU depicted in \Cref{fig:price_volume_uncertainty}).
  ]{\label{tab:mean_CVaR_with_volume_uncertainty}
  \begin{adjustbox}{minipage=\linewidth,scale=0.8}
  \centering
  \begin{tabular}[t]{|c|ccccc||c|}
      \hline
      &&&&&& \\[-12pt]
      $\lambda_{\CVaR}$ [1] & Expectation [\$] & $\CVaR_{0.1}$ [\$] & $\CVaR_{0.05}$ [\$] & $\CVaR_{0.01}$ [\$] &Variance [$\$^2$] & $\varphi_{0.1}^{\lambda_{\CVaR}}$ [\$] \\ [2pt]
      \hline
      0.00 & 2.973e5 & 1.353e6 & 1.589e6 & 2.096e6 & 3.096e11 & 2.973e5 \\
      0.25 & 3.331e5 & 1.084e6 & 1.268e6 & 1.676e6 & 1.464e11 & 5.210e5 \\
      0.50 & 4.015e5 & 9.638e5 & 1.113e6 & 1.456e6 & 7.877e10 & 6.826e5 \\
      0.75 & 4.586e5 & 9.274e5 & 1.058e6 & 1.366e6 & 5.474e10 & 8.102e5 \\
      1.00 & 4.982e5 & 9.211e5 & 1.041e6 & 1.330e6 & 4.502e10 & 9.211e5 \\
      \hline
    \end{tabular}
  \end{adjustbox}
  }
\end{table}

%% file: sections/3.3-framework_comparison.tex
\subsection{Comparison of the performance of the different frameworks}%
\label{sub:frameworks_comparison_numerical_experience}

In the previous section,
we noted that a trader who correctly assesses the volume uncertainty achieves better $\meanCVaR_{0.1}$ trade-offs at the expense of a greater variance of their trading costs.
In this section,
we compare our model with the ones in the literature.
\comreft{
  To evaluate the relative performance of each framework,
  we decided to use as comparison criteria the expectation and the $\alpha$-Conditional Value-at-Risk of the trading costs;
  the variance being not of utmost importance as justified hereafter.
  Additionally,
  we also compare the probability density function (PDF) and cumulative distribution function (CDF) of the trading costs of the different frameworks.
}

\paragraph{Mean-variance with recourse.}%
\label{par:mean_variance_with_recourse}

\begin{figure}
  \centering
  \hspace*{\fill}
  \subfloat{\label{fig:comparison_model_cdf}\includegraphics[width=0.47\linewidth]{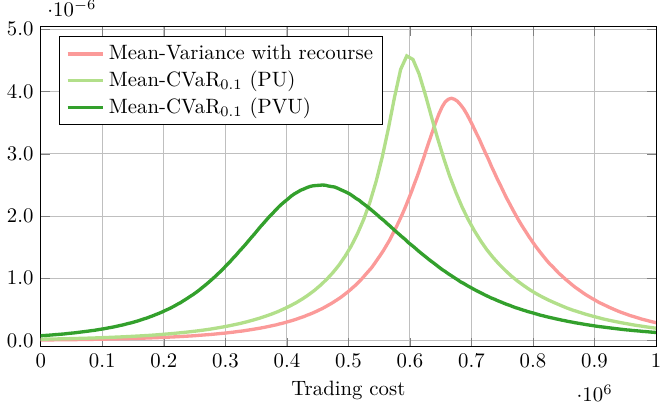}}
  \hspace*{\fill}
  \subfloat{\label{fig:comparison_model_pdf}\includegraphics[width=0.47\linewidth]{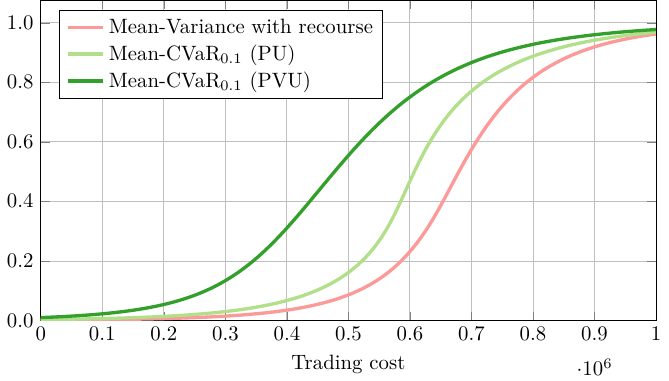}}
  \hspace*{\fill}
  \caption{
    Comparison of the PDFs and CDFs of the trading cost of a risk-averse trader computing their strategy
    with the mean-variance framework with recourse for $\lambda_{\Var} \to \infty$
    and with the mean-$\CVaR_{0.1}$ (PU and PVU) framework (Model~\eqref{eq:mean_CVaR_optimisation_problem}) for $\lambda_{\CVaR} = 1$
    (a number of $10^8$ random paths were generated to simulate these empirical distributions).
  }
  \label{fig:comparison_mean_variance_recourse_mean_CVaR_densities}
\end{figure}

First,
as mentioned previously,
the mean-variance framework of \cite{Almgren2001} is in practice not directly applicable in the presence of an uncertain volume target due to the constraint of Optimisation Problem~\eqref{eq:mean_variance_optimisation_problem}.
However,
a natural solution to circumvent this issue is to use it in conjunction with a systematic recourse at every trading period.
Since the objective function in the mean-variance and mean-$\CVaR_{\alpha}$ frameworks differs, 
it is not straightforward to compare the performance of the two models as one does not have a mapping between $\lambda_{\Var}$ and $\lambda_{\CVaR}$.
However,
the optimal strategy in the mean-variance framework converges when the risk-aversion increases.
Hence,
the parameter values $\lambda_{\CVaR} = 1$ and $\lambda_{\Var} \to +\infty$ both correspond to a trader that is exclusively risk focused;
these values enable us to compare the performance of both models.
The strategy of such a risk focused trader in the mean-variance framework with recourse consists of trading the entire initial forecast of the total volume to trade during the first trading period and then,
during the subsequent trading periods~$\tau_{i}$,
$i \in \dset{2}{m}$,
trading the amount $\rupdate_{i-1}$ of shares,
which corresponds to the forecast update unveiled at time $t_{i-1}$.
This strategy can be reproduced in our model by choosing $\ystratb = \pac{1, 0, \dots, 0}^{\T}$ and $\betab = \pa{\beta_{k,i}}$ with $\beta_{k,i} = 1$ if $i = k + 1$ and~0~otherwise.
The last line of \Cref{tab:mean_CVaR_without_volume_uncertainty,tab:mean_variance_with_recourse} reports the performance of an exclusively risk focused trader in both models;
note that in both situations the trader integrates only price uncertainty into their estimate of the trading cost but the actual trading costs are evaluated based on a market subjected to both sources of uncertainty.
The last line of \Cref{tab:mean_CVaR_with_volume_uncertainty},
which uses the same value of $\lambda_{\CVaR}$,
illustrates the benefit of factoring in the trade volume uncertainty:
both the expectation and the $\CVaR_{0.1}$ of the trading cost are significantly reduced.
The standard deviation of the trading cost slightly increases from 1.773e5 to 2.122e5,
but this is irrelevant:
Indeed,
\Cref{fig:comparison_mean_variance_recourse_mean_CVaR_densities},
which depicts the probability densities of the trading cost for both models,
shows empirically that the random variable that represents the trading cost in the mean-variance framework with recourse is first order stochastically dominant over the trading cost of the strategy that results from Model~\eqref{eq:mean_CVaR_optimisation_problem}.
As a consequence, 
a trader who is driven by profit should adopt the strategy proposed by Model~\eqref{eq:mean_CVaR_optimisation_problem} as the increased variance is due to increased downward risk,
which is beneficial.

\begin{table}[h]
  \caption{
    Performance of the strategies derived from the mean-variance framework with recourse for different risk-aversion parameter values $\lambda_{\Var}$.
  }
  \label{tab:mean_variance_with_recourse}
  \centering
  \begin{adjustbox}{minipage=\linewidth,scale=0.8}
    \centering
    \begin{tabular}[t]{|c|ccccc|}
      \hline
      &&&&&\\[-12pt]
      $\lambda_{\Var}$ [1/\$] & Expectation [\$] & $\CVaR_{0.1}$ [\$] & $\CVaR_{0.05}$ [\$] & $\CVaR_{0.01}$ [\$] & Variance [$\$^2$]  \\[1pt]
      \hline
      0.0            & 2.974e5 & 1.351e6 & 1.587e6 & 2.094e6 & 3.087e11 \\
      $10^{-6}$      & 3.626e5 & 1.016e6 & 1.182e6 & 1.554e6 & 1.076e11 \\
      $10^{-5}$      & 5.616e5 & 9.336e5 & 1.042e6 & 1.302e6 & 3.335e10 \\
      $10^{-4}$      & 6.716e5 & 9.935e5 & 1.088e6 & 1.319e6 & 2.722e10 \\
      $\to + \infty$ & 6.912e5 & 1.008e6 & 1.102e6 & 1.328e6 & 2.698e10 \\
      \hline
    \end{tabular}
  \end{adjustbox}
\end{table}

\paragraph{\comrefo{Exponential utility and mean-QV.}}%
\label{par:_utility_maximisation_and_mean_qv_}

We have previously seen that the model proposed by \citet{Cheng2017} could be slightly transformed in order to tackle a similar problem to ours but in continuous time.
In their model,
the authors assume that a risk-averse trader maximises
either the exponential utility of their final \pnl (profit-and-loss)  $\widetilde{\Pi}_{T}\pa{z}$,
or the expectation of the final \pnl penalised by its quadratic variation.
In both situations,
the authors provide closed form expressions for the optimal trading trajectory $z_{t}^{\star}$ when no residual volume is tolerated,
\ie when $\beta \to \infty$.
When allowing enough trading periods,
\eg $m = 10^2$,
the discrete approximation of the strategy derived from the model of \citet{Cheng2017} has nearly exactly the same final \pnl compared to the one of the continuous strategy.
Hence,
to compare their model with ours,
let assume the situation (a) described in \Cref{tab:market_parameters} with the difference that we assume one trading day ($T = 1$) divided in one hundred trading periods ($m = 10^2$).
Note that the final \pnl $\widetilde{\Pi}_{T}\pa{z}$ of a strategy $z$ is equivalent to minus its trading cost.

\begin{table}[h]
  \centering
  \caption{
  Trading cost comparison of the trading strategies for the different frameworks studied with an execution period subdivided in $m = 10^2$ trading periods.}
  \label{tab:perf_comparison}
  \subfloat[
  Mean-variance with recourse
  ]{\label{tab:perf_comparison_mean_variance_recourse}
    { \footnotesize
      \begin{tabular}{|C{1.5cm}|ccccc|}
	\hline
	&&&&&\\[-10pt]
	$\lambda_{\Var}$ [1/\$] & Expectation [\$] & $\CVaR_{0.1}$ [\$] & $\CVaR_{0.05}$ [\$] & $\CVaR_{0.01}$ [\$] & Variance [$\$^2$]  \\[1pt] 
	\hline
	0               & 8.133e5 & 1.411e6 & 1.532e6 & 1.784e6 & 1.006e11 \\
	$10^{-6}$       & 8.154e5 & 1.399e6 & 1.519e6 & 1.765e6 & 9.557e10 \\
	$10^{-5}$       & 9.155e5 & 1.409e6 & 1.510e6 & 1.722e6 & 6.792e10 \\
	$10^{-4}$       & 2.071e6 & 2.424e6 & 2.496e6 & 2.647e6 & 3.547e10 \\
	$\to + \infty$  & 6.311e7 & 6.330e7 & 6.334e7 & 6.341e7 & 9.803e9 \\
	\hline
      \end{tabular}
    }
  }\\
  \subfloat[
  Exponential utility
  ]{\label{tab:perf_comparison_exponential_utility}
    { \footnotesize
      \begin{tabular}{|C{1.5cm}|ccccc|}
	\hline
	&&&&&\\[-10pt]
	$\theta_{\EU}$ [1] & Expectation [\$] & $\CVaR_{0.1}$ [\$] & $\CVaR_{0.05}$ [\$] & $\CVaR_{0.01}$ [\$] & Variance [$\$^2$]  \\[1pt]
	\hline
	$  \to 0 $         & 7.861e5 & 1.264e6 & 1.348e6 & 1.512e6 & 7.423e10 \\
	$2 \cdot 10^{-6}$  & 7.883e5 & 1.252e6 & 1.334e6 & 1.493e6 & 7.002e10 \\
	$2 \cdot 10^{-5}$  & 9.354e5 & 1.313e6 & 1.379e6 & 1.510e6 & 4.631e10 \\
	$2 \cdot 10^{-4}$  & 3.466e6 & 3.640e6 & 3.670e6 & 3.730e6 & 9.814e9 \\
	$  \to + \infty $  & 2.780e7 & 2.794e7 & 2.797e7 & 2.802e7 & 7.081e9 \\
	\hline
      \end{tabular}
    }
  }\\
  \subfloat[
  Mean-QV
  ]{\label{tab:perf_comparison_mean_qv}
    { \footnotesize
      \begin{tabular}{|C{1.5cm}|ccccc|}
	\hline
	&&&&&\\[-10pt]
	$\lambda_{\QV} [1/\$]$ & Expectation [\$] & $\CVaR_{0.1}$ [\$] & $\CVaR_{0.05}$ [\$] & $\CVaR_{0.01}$ [\$] & Variance [$\$^2$]  \\[1pt]
	\hline
	$  \to 0 $       & 7.860e5 & 1.263e6 & 1.346e6 & 1.508e6 & 7.389e10 \\
	$10^{-6}$        & 7.885e5 & 1.255e6 & 1.336e6 & 1.495e6 & 7.075e10 \\
	$10^{-5}$        & 8.791e5 & 1.276e6 & 1.346e6 & 1.482e6 & 5.133e10 \\
	$10^{-4}$        & 1.637e6 & 1.902e6 & 1.948e6 & 2.038e6 & 2.282e10 \\
	$ \to + \infty $ & 2.534e6 & 2.741e6 & 2.777e6 & 2.848e6 & 1.393e10 \\
	\hline
      \end{tabular}
    }
  }\\
  \subfloat[
  Mean-$\CVaR_{0.1}$ (PVU)
  ]{\label{tab:perf_comparison_mean_cvar_pvu}
    { \footnotesize
      \begin{tabular}{|C{1.5cm}|ccccc|}
	\hline
	&&&&&\\[-10pt]
	$\lambda_{\CVaR}$ [1] & Expectation [\$] & $\CVaR_{0.1}$ [\$] & $\CVaR_{0.05}$ [\$] & $\CVaR_{0.01}$ [\$] & Variance [$\$^2$] \\[1pt]
	\hline
	0.00 & 8.132e5 & 1.409e6 & 1.530e6 & 1.780e6 & 1.001e11 \\
	0.25 & 8.153e5 & 1.398e6 & 1.517e6 & 1.764e6 & 9.529e10 \\
	0.50 & 8.206e5 & 1.387e6 & 1.503e6 & 1.742e6 & 9.035e10 \\
	0.75 & 8.286e5 & 1.382e6 & 1.495e6 & 1.729e6 & 8.597e10 \\
	1.00 & 8.396e5 & 1.380e6 & 1.491e6 & 1.722e6 & 8.186e10 \\
	\hline
      \end{tabular}
    }
  }
\end{table}

\citet{Cheng2017} first argue that the strategy resulting from the exponential utility (EU) maximisation can be regarded as a version of an adaptive Almgren-Chriss strategy.
When the uncertainty of the volume target disappears,
\ie $m_{0} \to 0$,
the optimal strategy given the risk-aversion parameter~$\theta_{\EU}$ obtained via the exponential utility framework recovers the classical Almgren-Chriss strategy with parameter value  $\lambda_{\Var}~\!\!=~\!\!\frac{\theta_{\EU}}{2}$.
Hence,
one can consider that a risk-averse trader with risk-aversion parameter $\lambda_{\Var}$ in the mean-variance framework would have a risk-aversion parameter $\theta_{\EU} $ equal to $ 2\lambda_{\Var}$ in the exponential utility framework;
this provides a mapping between $\lambda_{\Var}$ and $\theta_{\EU}$ that allows the comparison of both frameworks across the range of risk-aversion levels.
By comparing \Cref{tab:perf_comparison_mean_variance_recourse,tab:perf_comparison_exponential_utility},
we observe that the exponential utility framework produces results of the same order as the mean-variance framework with recourse.

While \citet{Cheng2017} employ the exponential utility for the sake of tractability,
the authors also provide,
in the special case where the uncertainty of order fills is modelled with a diffusion process,
a closed form solution of the optimal trading rate in the mean-QV framework by solving analytically the dynamic program.
The authors motivate the use of the quadratic variation since,
as pointed out by \citet{Almgren2012},
it captures the bulk of uncertainty.
The optimal trading rate in this model can be regarded as an adaptive Almgren-Chriss strategy with the difference that the characteristic time scales of liquidation are slightly different due to the uncertain order fills \citep{Cheng2017}.
When $m_{0} \to 0$,
the optimal strategy given the risk-aversion parameter~$\lambda_{\QV}$ recovers the classical Almgren-Chriss strategy with parameter value $\lambda_{\Var}~\!\!=~\!\!\lambda_{\QV}$.
This provides a mapping between $\lambda_{\Var}$ and $\lambda_{\QV}$.
The performance of the strategies derived with this model given different risk-aversion parameter values $\lambda_{\QV}$ is given in \Cref{tab:perf_comparison_mean_qv}:
the more risk-averse the trader,
the smaller the variance of their trading cost.
Compared to the two previous models (mean-variance with recourse and exponential utility maximisation),
for large risk-aversion parameters we observe a gain of one order of magnitude for both the expectation and $\CVaR_{0.1}$ of the trading costs, whereas the variance stays approximately of the same order.

\begin{figure}[h]
  \centering
  \hspace*{\fill}
  \subfloat{\label{fig:comparison_model_pdf}\includegraphics[width=0.47\linewidth]{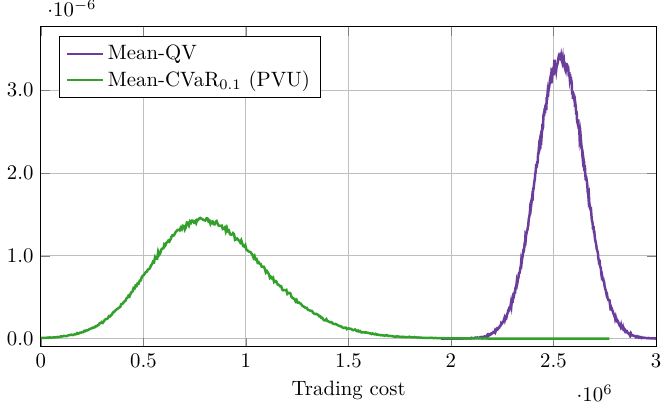}}
  \hspace*{\fill}
  \subfloat{\label{fig:comparison_model_cdf}\includegraphics[width=0.47\linewidth]{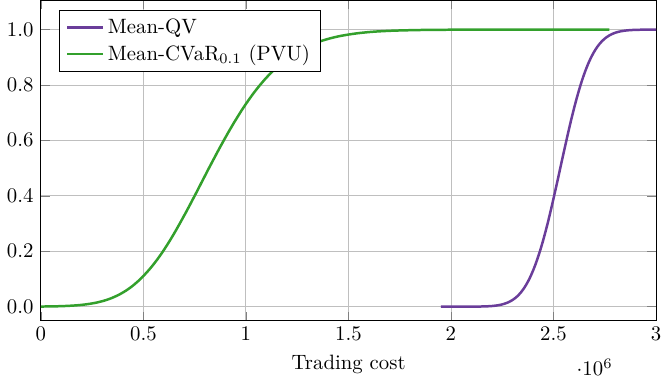}}
  \hspace*{\fill}\\
  \caption{
    Comparison of the PDFs and CDFs of the trading cost of a risk-averse trader computing their strategy
    with the mean-QV framework proposed by \citet{Cheng2017} for $\lambda_{\QV} \to \infty$
    and with the mean-$\CVaR_{0.1}$ (PVU) framework (Model~\eqref{eq:mean_CVaR_optimisation_problem}) for $\lambda_{\CVaR} = 1$
    (a number of $10^6$ random paths were generated to simulate these empirical distributions).
  }
  \label{fig:comparison_mean_qv_cvar_densities}
\end{figure}

Finally,
we consider the model presented in this paper where a trader considers the uncertainty arising from both the price dynamics and the updates on the volume target,
\ie the $\meanCVaR_{0.1}$ (PVU) framework.
The performance of the strategies obtained with this model is shown in \Cref{tab:perf_comparison_mean_cvar_pvu}.
In a first step,
as a sanity check,
let us consider a risk-neutral trader,
\ie
$\lambda_{\Var} = 0$,
$\theta_{\EU} \to 0$,
$\lambda_{\QV} = 0$,
and
$\lambda_{\CVaR} = 0$.
In this case all the frameworks minimise the same objective function,
\ie the expectation of the trading costs.
The first lines of \Cref{tab:perf_comparison_mean_variance_recourse,tab:perf_comparison_mean_cvar_pvu}
show that the mean-variance with recourse and the $\meanCVaR_{0.1}$ (PVU) frameworks produce nearly identical results;
this is coherent with \Cref{lemma:mean_CVaR_reproducibility_mean_variance_with_recourse}.
Indeed,
since the redistribution matrix $\betab$ was chosen as the one implied by the risk-neutral strategy of the mean-variance framework computed at time $t_{0}$ (see \Cref{eq:red_mat_num_res}),
the optimal strategy in the mean-variance framework with recourse belongs to the feasible set of strategies of the $\meanCVaR_{0.1}$ (PVU) framework.
Besides,
we observe that the exponential utility and mean-QV frameworks are performing better than the two previous frameworks.
Since both frameworks solve the same dynamical program,
the discrepancy between the values reported in \Cref{tab:perf_comparison_exponential_utility,tab:perf_comparison_mean_qv} for a risk-neutral trader can be attributed to numerical errors.
This better performance arises from the fact that these frameworks solve the dynamic program and reach consequently the optimal trading strategy.
We can therefore estimate the sub-optimality gap of our model to $\frac{\pa{8.129e5 - 7.861e5}}{7.855e5} \simeq 3.5\%$.
This can be seen as the price to pay of taking only partial recourse via a predetermined rule.
In a second step,
we compare our model with the mean-QV framework for an exclusively risk-focused trader,
which can be modelled by taking $\lambda_{\QV} \to +\infty$ and $\lambda_{\CVaR} =~\!\!1$.
\Cref{tab:perf_comparison_mean_qv,tab:perf_comparison_mean_cvar_pvu} illustrate that,
if a trader allows for a large variance of their trading cost,
then a significant reduction in the expectation and $\alpha$-Conditional Value-at-Risk of the trading costs can be achieved by using our new model compared to the mean-QV framework.
Similarly as for the comparison with the mean-variance framework with recourse,
we observe empirically in \Cref{fig:comparison_mean_qv_cvar_densities} that the random variable representing the trading cost in the mean-QV framework is first order stochastically dominant over the trading cost of the strategy resulting from Model~\eqref{eq:mean_CVaR_optimisation_problem}.
Again,
letting the variance increase enables our model to shift the distribution of the trading cost to smaller values,
which is beneficial.
In fact,
the increase in variance is due to a larger spread in lower trading costs.

\begin{figure}[h]
  \centering
  \hspace*{\fill}
  \subfloat{\label{fig:comparison_qv_pdf}\includegraphics[width=0.47\linewidth]{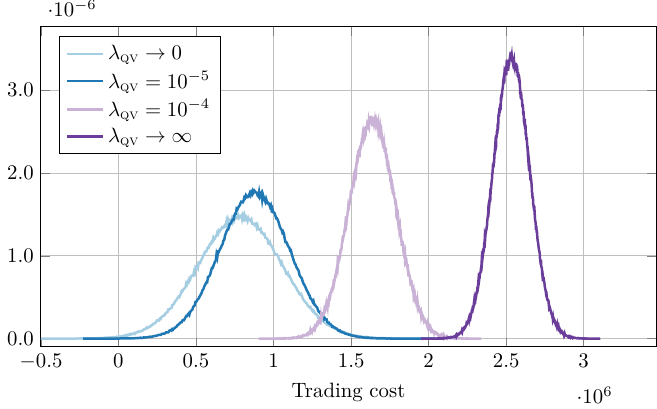}}
  \hspace*{\fill}
  \subfloat{\label{fig:comparison_qv_cdf}\includegraphics[width=0.47\linewidth]{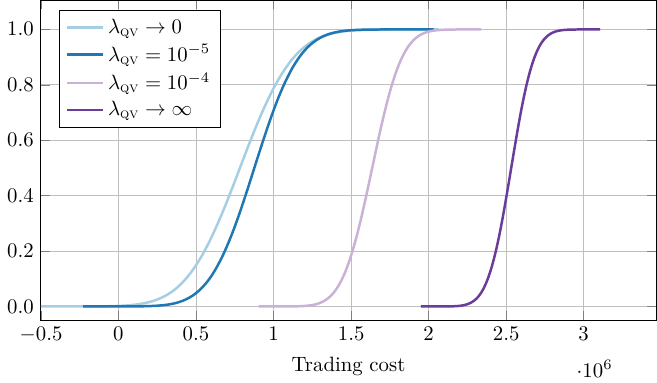}}
  \hspace*{\fill}\\
  \caption{
    Comparison of the PDFs and CDFs of the trading cost of a risk-averse trader computing their strategy
    with the mean-QV framework proposed by \citet{Cheng2017} for different risk-aversion parameter values
    (a number of $10^6$ random paths were generated to simulate these empirical distributions).
  }
  \label{fig:comparison_qv_densities}
\end{figure}

Finally,
\Cref{tab:perf_comparison} reports that a risk-neutral trader in the mean-QV framework achieves a smaller value for the $\CVaR_{0.1}$ of the trading costs compared to a trader that exclusively aims to minimise  $\CVaR_{0.1}$ in the $\meanCVaR_{0.1}$ (PVU) framework (\ie $\lambda_{\CVaR} = 1$).
This may call into question the advantage of our model,
however two important points should be brought to attention.
Firstly,
despite the apparent better performance of the mean-QV framework,
one should keep in mind that the analytical solution of the dynamic program to solve to obtain the optimal mean-QV trading rate is available only for special well behaved cases.
In the general case,
obtaining an analytical solution of the dynamic program would be more challenging,
if not impossible,
and obtaining a numerical solution would be significantly more computationally expensive compared to our approach.
Secondly,
the quadratic variation fails to correctly take into account the risk-aversion of a trader.
Indeed,
as represented in \Cref{fig:comparison_qv_densities} and reported in \Cref{tab:perf_comparison_mean_qv},
one roughly observes that,
the greater the risk-aversion level of the trader (\ie $\lambda_{\QV}$),
the greater the expectation and $\CVaR_{0.1}$ of their trading costs.
Hence,
the decrease of the variance of the trading costs is coupled with a shift of the distribution of the trading costs to greater values,
which is a rationale that would not be encountered in practice. 

In conclusion,
our model is significantly more versatile as it does not rely on any assumptions on the underlying distributions modelling the uncertainty related to the price dynamics and the forecast updates of the volume target.
It is also computationally cheaper than the alternative approaches,
and it coherently models the risk-aversion level of a trader:
the expectation of the worst-cases trading costs decreases as the risk-aversion level of the trader increases.


%% file: sections/3.4-power-forecast.tex
\subsection{{\comadd{Real world application: power trading}}}%
\label{sub:application_power}

One of the main advantages of our model is that it does not rely on any particular assumptions on the distributions of the price moves and forecast updates.
It can consequently be applied to a wider range of applications compared to the models of \citet{Almgren2001} and \citet{Cheng2017}.
Here,
we show how our model can be applied to power trading,
where for each half hourly settlement period,
power wholesalers (\resp producers) trade
power future contracts with the objective of achieving a contractual net position that matches the expected power demand of their end-consumers (\resp power generation).

In Great Britain (GB),
for every given settlement period,
the market participants can trade future contracts from several years ahead down until the market closure,
which occurs one hour before the start of delivery,
\ie the start of the settlement period.
For any given settlement period starting at time~$T$,
a wholesaler aims to achieve a final net position close to the realised power demand~$\rtarget_{T}$ of their end-consumers.
If they supply power to a fraction $\rho$ of the population,
the power demand of their customers is here assumed to be given by $\rtarget_{T} = \rho \rtargetsum_{T}$,
where $\rtargetsum_{T}$ denotes the national power demand during the settlement period.
Any residual volume resulting from the disparity between their contractual position against the realised demand $\rtarget_{T}$ is charged at the imbalance price,
which results from the balancing mechanism operated by the System Operator after the market closure.
For any half-hour settlement period,
there is consequently no obligation for a market participant to balance their contractual positions against the expected demand (or \resp generation).
However,
the exposure to this imbalance price drives the incentive to avoid any residual volume. 

We investigate here how a risk-averse power wholesaler in GB should acquire power positions ahead of a given settlement period in order to minimise their costs while factoring in risk as suggested by our Model~\eqref{eq:mean_CVaR_optimisation_problem}.
The market is modelled as follows:
The settlement period is one half-hour long and starts at the delivery time~$T$.
The trading window spans from 48 to 2 half-hours ahead of delivery time;
the initial trading time is denoted~$t_{0}$.
The trading window is subdivided in 46 one half-hour long trading periods.
To be in line with the notation introduced in \Cref{sub:mean_CVaR_framework},
the $i$-th trading period is denoted~$\tau_{i}$ and corresponds to the half-hour period that starts $48 - i + 1$ half-hours ahead of the settlement period (see \Cref{fig:time_line_power_trading}).

\begin{figure}[h]%
  \centering
  \resizebox{\linewidth}{!}{
    \input{\pathImagesTexOptimalTrading/strategy_power_update_source_files/strategy_power_update_im3}
  }
  \caption{Power trading time line and modelling assumptions.}
  \label{fig:time_line_power_trading}
\end{figure}

\begin{figure}[h]
  \centering
  \hspace*{\fill}
  \subfloat[]{\label{fig:forecast_mean_abs_error}
    \includegraphics[width=0.47\linewidth]{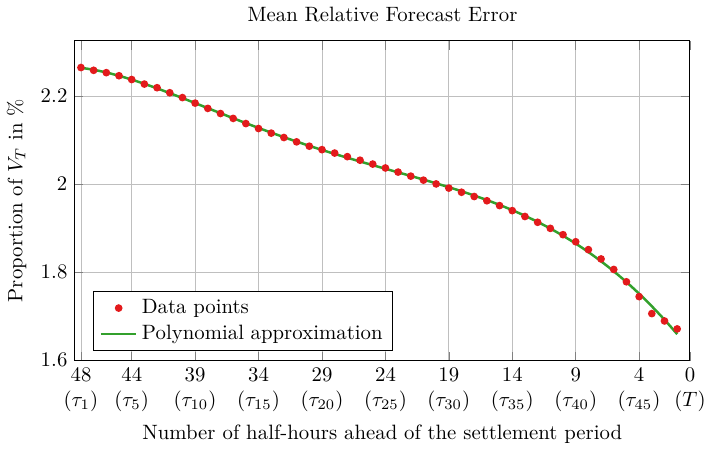}
  }
  \hspace*{\fill}
  \subfloat[]{\label{fig:forecast_prob_update}
    \includegraphics[width=0.47\linewidth]{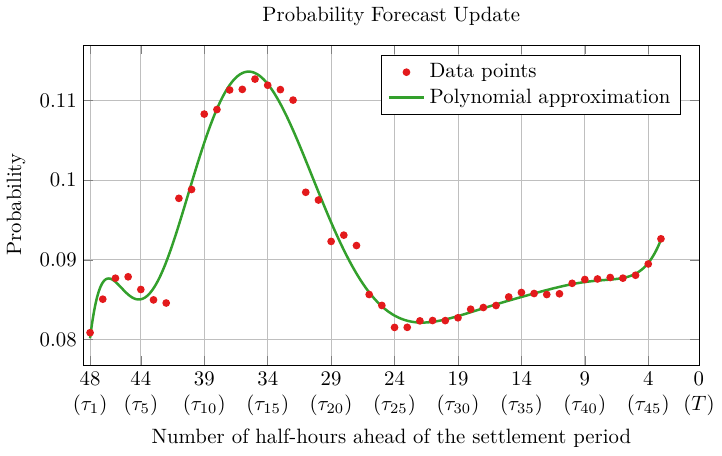}
  }
  \hspace*{\fill}\\
  \hspace*{\fill}
  \subfloat[]{\label{fig:forecast_mean_update}
    \includegraphics[width=0.47\linewidth]{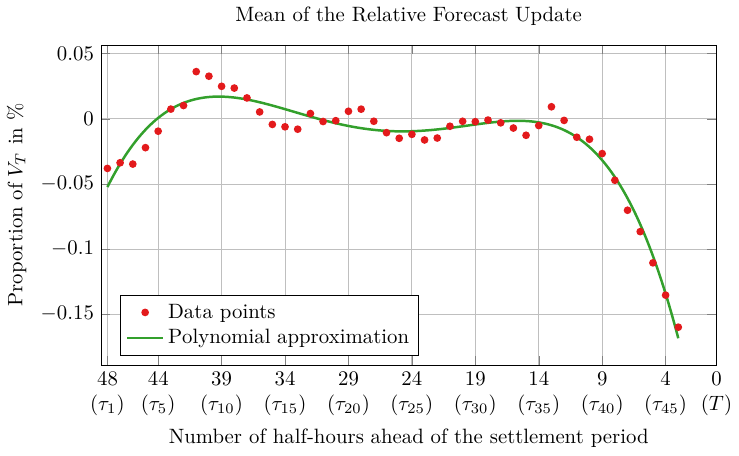}
  }
  \hspace*{\fill}
  \subfloat[]{\label{fig:forecast_std_update}
    \includegraphics[width=0.47\linewidth]{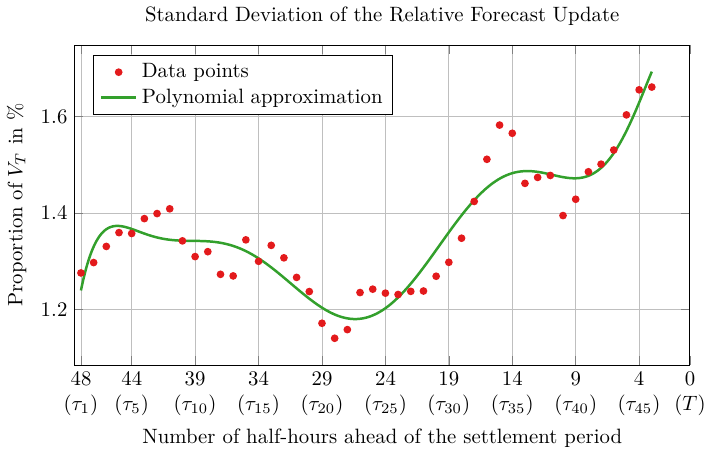}
  }
  \hspace*{\fill}
  \caption{
    Analysis of the forecast updates of the national power demand in GB.
  }
  \label{fig:analysis_demand_forecast}
\end{figure}

To make their trading decisions,
the market participants use the operational data relating to the GB electricity balancing and settlement arrangements that is published by the \emph{Balancing Mechanism Reporting Service} (BMRS).
Using the data of the past five years,
\Cref{fig:analysis_demand_forecast} represents the trend of the forecast updates of the national power demand in GB:
\Cref{fig:forecast_mean_abs_error} illustrates the mean of the relative forecast error,
which is computed as follows
$\abs{\rtargetsum_{i} - \rtargetsum_{T}}/\rtargetsum_{T}$,
where $\rtargetsum_{T}$ denotes the realised national demand and $\rtargetsum_{i}$ the demand forecast unveiled after the $i$-th trading period~$\tau_{i}$,
\ie at decision time~$t_{i}$.
\Cref{fig:forecast_prob_update} represents the probability $p_{i}$ of having a forecast update of the power demand at time~$t_{i}$.
Conditional on there being a forecast update,
\Cref{fig:forecast_mean_update,fig:forecast_std_update} represent respectively
the mean $m_{i}$ and standard deviation $s_{i}$ of the relative forecast update,
which is given by
$\pa{ \rtargetsum_{i} - \rtargetsum_{i-1} } / \rtargetsum_{T}$.
For all graphs,
we have fitted a polynomial approximation of the data points;
let~$\hat{p}_{i}$,
$\hat{m}_{i}$,
and~$\hat{s}_{i}$ denote respectively the value of the polynomial approximating the probability,
the mean and the standard deviation of the forecast update that comes to the trader's knowledge at time~$t_{i}$.
Based on these values,
we assume that,
for each trading period~$\tau_{i}$ with~$i \le 45$,
a wholesaler estimates the distribution of the related forecast update on the power demand of their customers as follows:
Firstly,
there is a probability~$\hat{p}_{i}$ of having a forecast update.
Secondly,
conditional on a forecast update occurring,
the forecast update~$\rupdate_{i}$ is assumed to follow the normal distribution
$\normald{\hat{m}_{i}\bar{\rtarget}_{T}}{\hat{s}_{i}^{2}\bar{\rtarget}_{T}^{2}}$,
where $\bar{\rtarget}_{T} = \rho \bar{\rtargetsum}_{T}$ and $\bar{\rtargetsum}_{T}$ denotes the average realised national power demand.
Finally,
because the market closure happens one hour before the start of delivery,
a trader cannot take into consideration in his trading strategy any forecast update released after the decision time~$t_{45}$.
Therefore,
the residual volume that should have been traded and that will be exchanged at the imbalance price is given by the random variable
$\rho \pa{\rtargetsum_{T} - \rtargetsum_{45}}$.
This is taken into consideration in our model by assuming that the forecast update~$\rupdate_{46}$ of the last trading period follows the normal distribution
$\normald{m_{45}^{err}\bar{\rtarget}_{T}}{\pa{s_{45}^{err}}^{2}\bar{\rtarget}_{T}^{2}}$,
where~$m_{45}^{err}$ and~$s_{45}^{err}$ are respectively the average and standard deviation of the relative forecast error at decision time~$t_{45}$,
\ie
$\pa{\rtargetsum_{T} - \rtargetsum_{45}} / \rtargetsum_{T}$.
Additionally to these 46 trading periods,
one adds a virtual 47-th trading period,
which corresponds to the balancing period after the market closure.
Note that during this balancing period,
we artificially assume that there is no forecast update on the demand as it has already been taken into consideration in the forecast update distribution of $\tau_{46}$.
To incentivise the trader to trade the entirety of their position before the balancing period $\tau_{47}$,
we penalise any outstanding volume by fixing a larger temporary impact parameter~$\eta_{47}$ for the balancing period $\tau_{47}$ compared to the other trading periods;
a value three times larger is here used (see \Cref{tab:power_market_parameters}).
In a similar vein as the penalty term parametrised by~$\beta$ in the model of \citet{Cheng2017},
the larger~$\eta_{47}$,
the more any outstanding volume is penalised.
As suggested in~\Cref{sub:frameworks_comparison_theory},
we assume that the residual volumes to trade due to the forecast updates are redistributed over the future trading periods based on the redistribution matrix~$\betab$ implied by the risk-neutral optimal trading strategy~$\ystratb^{\star, t_{0}, \lambda_{\Var} = 0}$ in the mean-variance framework.


From the perspective of the price distributions,
we consider an initial price~$\rprice_{0}$ of $90$.
For the sake of simplicity (the data not being publicly available),
we assume that the volatility $\sigma$ of the asset is constant during the entire execution period and is equal to $0.9$,
and that for each trading period $\tau_{i}$ the price shift~$\rpriceshift_{i}$ follows the normal distribution
$\normald{0}{\tau_{i}\sigma^{2}}$.
The rest of the parameters are fixed as follows:
the average realised national power demand $\rtargetsum_{T}$ equals $3 \cdot 10^5 \si{\mega \watt}$.
The wholesaler provides power to $\rho = \frac{1}{3}$ of the population,
has an initial position of 0,
and has an initial estimate of the power demand of the end-consumers equal to $\rtarget_{0} = \bar{\rtarget}_{T} = \rho \bar{\rtargetsum}_{T} =  10^5$.
Finally,
the bid-ask spread of each trading period is assumed to be equal to 0.25\% of the initial asset price $\rprice_{0}$,
\ie $\forall i \in \dseto{m}: b_{i} = 0.225$;
the values used for temporary and permanent impact parameters can be found in \Cref{tab:power_market_parameters}.

\begin{table}[h]
  \begin{center}
    {\footnotesize
      \caption{Power market parameters}
      \label{tab:power_market_parameters}
      \begin{tabular}{|rl|}
	\hline
	& \\[-8pt]
	Number of trading periods: & $m = 47$\\
	Trading period length:& $\forall i \in \dset{1}{m}: \tau_{i} = \frac{1}{48}$ [day]\\
	Initial future contract price:  &$\rprice_{0} = 90$ [£/share] \\
	Asset's volatility: & $\sigma = 0.9$ [\si{(£ \per share) \per day^{1/2}}] \\
	Price shifts distribution:& $\forall i \in \dset{1}{m}: \rpriceshift_{i} \sim \normald{0}{\tau_{i}\sigma^{2}}$\\
	Average realised national power demand: & $\bar{\rtargetsum}_{T} = 3 \cdot 10^5$ [\si{\mega \watt}]\\
	Proportion of the population supplying: & $\rho = \frac{1}{3}$ \\
	Initial forecast of the customers' power demand: & $\rtarget_{0} = \bar{\rtarget}_{T} = \rho \bar{\rtargetsum}_{T} =  10^5$ [\si{\mega \watt}]\\
	Volume forecast updates distribution:& $\forall i \in \dset{1}{45}: \rupdate_{i} \sim \normald{\hat{m}_{i}\bar{\rtarget}_{T}}{\hat{s}_{i}^{2}\bar{\rtarget}_{T}^{2}}$\\ & $\rupdate_{46} \sim \normald{m_{45}^{err}\bar{\rtarget}_{T}}{\pa{s_{45}^{err}}^{2}\bar{\rtarget}_{T}^{2}}$ \\ & $\rupdate_{47} = 0$ \\
	Bid-ask spread: & $\forall i \in \dset{1}{m}: b_{i} = 0.25\% \, \rprice_{0} = 0.225$ [£/share]\\
	Fixed cost: & $\forall i \in \dset{1}{m}: \epsilon_{i} = 0.5b_{i} = 0.1125$ [£/share]\\
	Impact at 2\% of market: & $\forall i \in \dset{1}{46}: \eta_{i} = b_{i}/(0.02\cdot \rtargetsum_{T}) = 3.75 \cdot 10^{-5}$ [\si{(£ \per share) \per (share \per day)}]\\ & $\eta_{47} = 3 \cdot b_{47}/(0.02\cdot \rtargetsum_{T}) = 11.25 \cdot 10^{-5}$ [\si{(£ \per share) \per (share \per day)}]\\
	Permanent impact parameter:& $\forall i \in \dset{1}{m}: \gamma_{i} = (5 b_{i})/(0.1\rtargetsum_{T}) = 3.75 \cdot 10^{-5}$ [\si{£ \per share^2}]\\
	$\CVaR_\alpha$ parameter:& $\alpha = 0.1$ \\[4pt]
	\hline
      \end{tabular}
    }
  \end{center}
\end{table}


Similarly to \Cref{fig:mean_CVaR_price_and_volume_uncertainty},
\Cref{fig:power_pvu} depicts the trading strategies returned by our model for a risk-averse wholesaler in GB when they consider the uncertainty related to both the price dynamics and the power demand of the end-consumers.
For different risk-aversion levels,
\Cref{fig:power_diff_pvu_pu} depicts how the trading strategy is modified compared to the situation where the wholesalers wrongly assumes that the power demand of the end-users is perfectly known at the start of the execution period.
The difference in performance of the trading strategies obtained in both situations is listed in \Cref{tab:power_performance}.
When one considers the uncertainty related to the power demand,
better mean-$\CVaR_{0.1}$ trade-offs are achieved for any risk-level.
We do not further analyse the results as similar comments can be made compared to the test cases previously presented (see \Cref{sub:optimal_strategies,sub:model_performance}). 

\begin{figure}[h]
  \centering
  \hspace*{\fill}
  \subfloat[]{\label{fig:power_pvu}
    \resizebox{!}{0.2\textheight}{
      \begin{tikzpicture}[]
	\begin{groupplot}[
	  group style={
	    group name=my fancy plots,
	    group size=1 by 2,
	    xticklabels at=edge bottom,
	    vertical sep=0pt
	  },
	  xtick={1,5,10,15,20,25,30,35,40,47},
	  xticklabels={$\tau_{1}$,$\tau_{5}$,$\tau_{10}$,$\tau_{15}$,$\tau_{20}$,$\tau_{25}$,$\tau_{30}$,$\tau_{35}$,$\tau_{40}$,$\tau_{47}$},
	  ]

	  \nextgroupplot[
	  ymin = 0.018, ymax = 0.025,
	  xmin = 0, xmax = 48,
	  ytick={0.02,0.022,0.024},
	  yticklabels={0.02,0.022,0.024},
	  xmajorgrids, ymajorgrids,
	  scaled ticks=false,
	  yticklabel style={
	    /pgf/number format/fixed,
	    /pgf/number format/precision=5
	  },
	  y label style={at={(-0.12,0.4)}},
	  axis x line=top, 
	  axis y discontinuity=parallel,
	  height=5.5cm,
	  ylabel = Proportion to trade,
	  xtick style={draw opacity=0},
	  ] 

	  \addplot table [x=T,y=Y,select coords between index={0}{45}] {\pathTxtFilesOptimalTrading/mean-cvar-optim_8873500018088811927.txt};
	  \addplot table [x=T,y=Y,select coords between index={0}{45}] {\pathTxtFilesOptimalTrading/mean-cvar-optim_3532228027801323971.txt};
	  \addplot table [x=T,y=Y,select coords between index={0}{45}] {\pathTxtFilesOptimalTrading/mean-cvar-optim_90573412915485638.txt};
	  \addplot table [x=T,y=Y,select coords between index={0}{45}] {\pathTxtFilesOptimalTrading/mean-cvar-optim_15497166153318989772.txt};
	  \addplot table [x=T,y=Y,select coords between index={0}{45}] {\pathTxtFilesOptimalTrading/mean-cvar-optim_17698336453125757495.txt};

	  \nextgroupplot[
	  ymin = -0.002, ymax = 0.012,
	  xmin = 0, xmax = 48,
	  xmajorgrids, ymajorgrids,
	  ytick={0.0,0.005,0.01},
	  yticklabels={0.0,0.005,0.01},
	  scaled ticks=false,
	  yticklabel style={
	    /pgf/number format/fixed,
	    /pgf/number format/precision=5
	  },
	  legend pos=south west,
	  legend entries={
	    $\lambda_{\CVaR} = 0.00$,
	    $\lambda_{\CVaR} = 0.25$,
	    $\lambda_{\CVaR} = 0.50$,
	    $\lambda_{\CVaR} = 0.75$,
	    $\lambda_{\CVaR} = 1.00$,
	  },
	  legend style={at={(0.05,1.0)},anchor=west},
	  axis x line=bottom,
	  height=3.0cm,
	  xlabel = Trading period,
	  ]

	  \addplot table [x=T,y=Y,select coords between index={46}{46}] {\pathTxtFilesOptimalTrading/mean-cvar-optim_8873500018088811927.txt};
	  \addplot table [x=T,y=Y,select coords between index={46}{46}] {\pathTxtFilesOptimalTrading/mean-cvar-optim_3532228027801323971.txt};
	  \addplot table [x=T,y=Y,select coords between index={46}{46}] {\pathTxtFilesOptimalTrading/mean-cvar-optim_90573412915485638.txt};
	  \addplot table [x=T,y=Y,select coords between index={46}{46}] {\pathTxtFilesOptimalTrading/mean-cvar-optim_15497166153318989772.txt};
	  \addplot table [x=T,y=Y,select coords between index={46}{46}] {\pathTxtFilesOptimalTrading/mean-cvar-optim_17698336453125757495.txt};

	\end{groupplot}
      \end{tikzpicture}
    }
  }
  \hspace*{\fill}
  \subfloat[]{\label{fig:power_diff_pvu_pu}
    \resizebox{!}{0.2\textheight}{
      \begin{tikzpicture}[]
	\begin{groupplot}[
	  group style={
	    group name=my fancy plots,
	    group size=1 by 3,
	    xticklabels at=edge bottom,
	    vertical sep=0pt
	  },
	  xtick={1,5,10,15,20,25,30,35,40,47},
	  xticklabels={$\tau_{1}$,$\tau_{5}$,$\tau_{10}$,$\tau_{15}$,$\tau_{20}$,$\tau_{25}$,$\tau_{30}$,$\tau_{35}$,$\tau_{40}$,$\tau_{47}$},
	  xmajorgrids,
	  ymajorgrids,
	  ytick pos=left,
	  ]

	  \nextgroupplot[
	  separate axis lines,
	  axis y discontinuity=parallel,
	  axis x line=top, 
	  height=3cm,
	  ymin = 0.0013,
	  ymax = 0.0017,
	  xmin = 0, xmax = 48,
	  ytick={0.0015,0.0016},
	  yticklabels={0.0015,0.0016},
	  xtick style={draw opacity=0},
	  scaled ticks=false,
	  yticklabel style={
	    /pgf/number format/fixed,
	    /pgf/number format/precision=5
	  },
	  ]

	  \addplot table [x=T,y=Y,select coords between index={46}{46}] {\pathTxtFilesOptimalTrading/difference_mean-cvar-optim_12345592210445333749_mean-cvar-optim_8873500018088811927.txt};

	  \nextgroupplot[
	  ymin = -0.0007, ymax = 0.0007,
	  xmin = 0, xmax = 48,
	  ytick={-0.0005,-0.00025,0.0,0.00025,0.0005},
	  yticklabels={-0.0005,-0.00025,0.0,0.00025,0.0005},
	  scaled ticks=false,
	  yticklabel style={
	    /pgf/number format/fixed,
	    /pgf/number format/precision=5
	  },
	  separate axis lines,
	  x axis line style={draw opacity=0},
	  xtick style={draw opacity=0},
	  height=5.5cm,
	  y label style={at={(-0.15,0.5)}},
	  ylabel = Proportion to trade,
	  axis y discontinuity=parallel,
	  ] 

	  \addplot table [x=T,y=Y,select coords between index={0}{45}] {\pathTxtFilesOptimalTrading/difference_mean-cvar-optim_12345592210445333749_mean-cvar-optim_8873500018088811927.txt};
	  \addplot table [x=T,y=Y,select coords between index={0}{45}] {\pathTxtFilesOptimalTrading/difference_mean-cvar-optim_7754849124706072148_mean-cvar-optim_3532228027801323971.txt};
	  \addplot table [x=T,y=Y,select coords between index={0}{45}] {\pathTxtFilesOptimalTrading/difference_mean-cvar-optim_12944883113089842232_mean-cvar-optim_90573412915485638.txt};
	  \addplot table [x=T,y=Y,select coords between index={0}{45}] {\pathTxtFilesOptimalTrading/difference_mean-cvar-optim_5114899910098603249_mean-cvar-optim_15497166153318989772.txt};
	  \addplot table [x=T,y=Y,select coords between index={0}{45}] {\pathTxtFilesOptimalTrading/difference_mean-cvar-optim_549416532610642337_mean-cvar-optim_17698336453125757495.txt};

	  \nextgroupplot[
	  ymin = -0.007, ymax = -0.001,
	  xmin = 0, xmax = 48,
	  ytick={-0.002,-0.004,-0.006},
	  yticklabels={-0.002,-0.004,-0.006},
	  scaled ticks=false,
	  yticklabel style={
	    /pgf/number format/fixed,
	    /pgf/number format/precision=5
	  },
	  legend pos=south west,
	  legend entries={
	    $\lambda_{\CVaR} = 0.00$,
	    $\lambda_{\CVaR} = 0.25$,
	    $\lambda_{\CVaR} = 0.50$,
	    $\lambda_{\CVaR} = 0.75$,
	    $\lambda_{\CVaR} = 1.00$,
	  },
	  legend style={at={(0.04,3.8)},anchor=west},
	  axis x line=bottom,
	  height=3.0cm,
	  xlabel = Trading period,
	  ]

	  \addplot table [x=T,y=Y,select coords between index={46}{46}] {\pathTxtFilesOptimalTrading/difference_mean-cvar-optim_12345592210445333749_mean-cvar-optim_8873500018088811927.txt};
	  \addplot table [x=T,y=Y,select coords between index={46}{46}] {\pathTxtFilesOptimalTrading/difference_mean-cvar-optim_7754849124706072148_mean-cvar-optim_3532228027801323971.txt};
	  \addplot table [x=T,y=Y,select coords between index={46}{46}] {\pathTxtFilesOptimalTrading/difference_mean-cvar-optim_12944883113089842232_mean-cvar-optim_90573412915485638.txt};
	  \addplot table [x=T,y=Y,select coords between index={46}{46}] {\pathTxtFilesOptimalTrading/difference_mean-cvar-optim_5114899910098603249_mean-cvar-optim_15497166153318989772.txt};
	  \addplot table [x=T,y=Y,select coords between index={46}{46}] {\pathTxtFilesOptimalTrading/difference_mean-cvar-optim_549416532610642337_mean-cvar-optim_17698336453125757495.txt};

	\end{groupplot}
      \end{tikzpicture}
    }
  }
  \hspace*{\fill}
  \caption{
    Trading power in GB:
    \Cref{fig:power_pvu}:
    optimal trading strategies under PVU,
    \Cref{fig:power_diff_pvu_pu}:
    difference between the strategies under PVU and PU.
  }
  \label{fig:power_strategies}
\end{figure}

\begin{table}[h]
  \caption{
    Trading power in GB:
    performance of the strategies obtained with~\eqref{eq:mean_CVaR_optimisation_problem} for different risk-aversion parameter values $\lambda_{\CVaR}$.
  }
  \label{tab:power_performance}
  \centering
  \subfloat[
  A trader considers that the only source of uncertainty is the price dynamics.
  ]{\label{tab:power_mean_CVaR_without_volume_uncertainty}
    \begin{adjustbox}{minipage=\linewidth,scale=0.8}
      \centering
      \begin{tabular}[t]{|c|ccccc||c|}
	\hline
      &&&&&& \\[-12pt]
      $\lambda_{\CVaR}$ [1] & Expectation [\$] & $\CVaR_{0.1}$ [\$] & $\CVaR_{0.05}$ [\$] & $\CVaR_{0.01}$ [\$] &Variance [$\$^2$] & $\varphi_{0.1}^{\lambda_{\CVaR}}$ [\$] \\[2pt]
      \hline
      0.00 & 6.102e5 & 7.572e5 & 7.945e5 & 8.841e5 & 5.558e9 & 6.102e5 \\
      0.25 & 6.103e5 & 7.566e5 & 7.938e5 & 8.832e5 & 5.503e9 & 6.469e5 \\
      0.50 & 6.105e5 & 7.562e5 & 7.933e5 & 8.825e5 & 5.449e9 & 6.834e5 \\
      0.75 & 6.109e5 & 7.559e5 & 7.929e5 & 8.819e5 & 5.396e9 & 7.196e5 \\
      1.00 & 6.114e5 & 7.559e5 & 7.928e5 & 8.818e5 & 5.345e9 & 7.559e5 \\
      \hline
    \end{tabular}
    \end{adjustbox}
  }\\
  \subfloat[
  A trader considers that the uncertainty arises from both the price dynamics and the power demand.
  ]{\label{tab:power_mean_CVaR_with_volume_uncertainty}
    \begin{adjustbox}{minipage=\linewidth,scale=0.8}
      \centering
      \begin{tabular}[t]{|c|ccccc||c|}
	\hline
      &&&&&& \\[-12pt]
      $\lambda_{\CVaR}$ [1] & Expectation [\$] & $\CVaR_{0.1}$ [\$] & $\CVaR_{0.05}$ [\$] & $\CVaR_{0.01}$ [\$] &Variance [$\$^2$] & $\varphi_{0.1}^{\lambda_{\CVaR}}$ [\$] \\ [2pt]
      \hline
      0.00 & 6.101e5 & 7.606e5 & 7.993e5 & 8.922e5 & 5.751e9 & 6.101e5 \\
      0.25 & 6.110e5 & 7.531e5 & 7.885e5 & 8.737e5 & 5.282e9 & 6.465e5 \\
      0.50 & 6.122e5 & 7.510e5 & 7.850e5 & 8.668e5 & 5.094e9 & 6.816e5 \\
      0.75 & 6.134e5 & 7.503e5 & 7.837e5 & 8.638e5 & 4.988e9 & 7.161e5 \\
      1.00 & 6.144e5 & 7.502e5 & 7.832e5 & 8.623e5 & 4.914e9 & 7.502e5 \\
      \hline
    \end{tabular}
      \end{adjustbox}
    }
\end{table}

%% file: images/tikz/strategy_power_update_source_files/strategy_power_update_im3.tex
  \begin{tikzpicture}[domain=0:10,scale=0.8]
    \label{fig:time_line_power_trading}
    \draw[-] (-1,-0.1) -- (-1,0.1) node[above] {$t_0$};
    \draw[-] (-1,0.8) -- (-1,1.3) node[above] {\small{Initial trading time}};
    \draw[-,thick] (-2,0) -- (2,0);
    \draw[-,thick,dotted] (2,0) -- (3,0);
    \draw[-,thick] (3,0) -- (7,0);
    \draw[-,thick,dotted] (7,0) -- (8,0);
    \draw[->,thick] (8,0) -- (16,0) node[right] {$time$};
    \draw[-] (0.5,-0.1) -- (0.5,0.1) node[above] {$t_1$};
    \draw[<->] (-1,-0.2) -- (0.5,-0.2) node[midway,below] {$\tau_{1}$};
    \draw[-] (4.5,-0.1) -- (4.5,0.1) node[above] {$t_{i-1}$};
    \draw[-] (6,-0.1) -- (6,0.1) node[above] {$t_i$};
    \draw[<->,color=Paired-D] (4.5,-0.2) -- (6,-0.2) node[midway,below] {$\tau_{i}$};
    \draw[->,thick,color=Paired-D] (5.25,-0.8) -- (5.25,-1.3) -- (4.75,-1.3) node[left] {
      \small{
	$
	\begin{cases}
	  \rupdate_{i} = 0 & \text{with probability } 1 - \hat{p}_{i} \\
	  \rupdate_{i}
	  \sim
	  \normald{\hat{m_{i}}\bar{\rtarget}_{T}}{\hat{s}_{i}^{2}\bar{\rtarget}_{T}^{2}} & \text{with probability } \hat{p}_{i}
	\end{cases}
	$
      }
    };
    \draw[-] (9,-0.1) -- (9,0.1) node[above] {$t_{45}$};
    \draw[-] (10.5,-0.1) -- (10.5,0.1) node[above] {$t_{46}$};
    \draw[-] (10.5,0.8) -- (10.5,1.4) node[above] {\small{Market closure}};
    \draw[<->,color=Paired-J] (9,-0.2) -- (10.5,-0.2) node[midway,below] {$\tau_{46}$};
    \draw[->,thick,color=Paired-J] (9.75,-0.8) -- (9.75,-2.5) -- (9.25,-2.5) node[left] {
      \small{
	$
	\rupdate_{46}
	\sim
	\normald{m_{45}^{err}\bar{\rtarget}_{T}}{\pa{s_{45}^{err}}^{2}\bar{\rtarget}_{T}^{2}}
	$
      }
    };
    \draw[-] (13.5,-0.1) -- (13.5,0.1) node[above] {$T$};
    \draw[-] (13.5,0.8) -- (13.5,1.3) node[above] {\small{Delivery time}};
    \draw[-] (15,-0.1) -- (15,0.1);
    \draw[<->,color=Paired-I,text width = 1cm] (13.5,-0.2) -- (10.5,-0.2) node[midway,below] {\small{Virtual period}};
    \draw[->,thick,color=Paired-I] (12.0,-1.5) -- (12.0,-2.15) node[below] {
      $
      \rupdate_{47} = 0
      $
    };
    \draw[<->,color=Paired-B] (13.5,-0.2) -- (15,-0.2) node[midway,below,text width=1.2cm] {\small{Settlement period}};
  \end{tikzpicture}

%% file: sections/4-comments.tex
\section{Discussion}%
\label{sec:comments}

\comreft{
Our choice to use the $\alpha$-Conditional Value-at-Risk as risk measure in Model~\eqref{eq:mean_CVaR_optimisation_problem} was guided by its widespread use and ease of interpretation,
despite the controversy that this risk measure may lead to time inconsistent optimal policies: 
the policy obtained by optimising the model at an intermediate stage of the execution period is not necessarily equivalent to the continuation of the optimal policy computed at the initial time $t_{0}$ \citep{Shapiro2009,Rudloff2014}.
In~\Cref{sec:numerical_results} we saw however that due to the forward planning of recourse with a predetermined trade volume re-allocation rule,
our model competes favourably with established alternatives,
yielding lower trading cost in expectation and $\alpha$-Conditional Value-at-Risk.
Our merit function accounts for the trade-off between the partially conflicting goals of lowering the expectation and the $\alpha$-Conditional Value-at-Risk of trading costs and is computationally cheaper than either taking recourse under the mean-variance framework whenever forecast updates on the volume target become available or solving a dynamic program in order to optimise the mean-QV trade-off.
These advantages of our model outweigh the disadvantage that the time inconsistency of the Mean-$\CVaR_{\alpha}$ risk measure may lead to suboptimal strategies \citep{Rudloff2014}. 
}

\comreft{
The model proposed in this paper should thus be considered as an alternative to the traditional methods for its competitive  performance,
its ease of interpretation,
and for its relatively cheap computational cost.
If one desires to take full recourse in order to further improve the performance of our model,
one could replace the $\CVaR_{\alpha}$ term by the dynamic coherent risk measure \citep{Riedel2004} induced by the $\CVaR_{\alpha}$ risk measure.
\citet{Lin2015} explored such an approach in the context of the traditional trade execution problem under price uncertainty only.
An extension to the case with volume uncertainty would imply the applicability of the dynamic programming principle, 
which guarantees the time consistency of the optimal strategies but at the cost of a greater computational time. 
This would be a great extension of our work and is left for future research.
}



Among the parameters for our model we need to specify a redistribution matrix to decide how any residual trading volumes that become apparent after forecast updates are redistributed over future trading periods.
In the numerical section of this paper we chose this redistribution matrix according to trade volume proportions implied by the risk-neutral optimal trading strategy $\ystratb^{\star,t_{0},\lambda_{\Var} = 0}$ under the mean-variance framework of \citet{Almgren2001}.
This choice is not guaranteed to be optimal,
and one could optimize over the redistribution matrix as well.
Although we did not address this issue in the present work,
we believe it is of interest for future research.
Numerical simulations we carried out suggest that a mild additional cost savings can be achieved.

Finally, in this paper we assumed that the random variables that model forecast updates are identically distributed.
In practice this assumption is often not justified, such as in 
power markets, where empirical data show that the standard deviation of forecast updates is time dependent and follows a ``$\cap$'' pattern over the trading window: Forecast updates at long time horizons are relatively small, since they are mainly based on the seasonal trend. At shorter time scales,
\ie for time horizons ranging from a couple of weeks to several hours ahead of the delivery time,
the forecast updates are more significant,
since weather forecasts are available and steadily become more precise.
Finally, the last forecast updates represent slight adjustments to ensure that the market clearing constraint is satisfied. 
\comreft{
Another example where inhomogeneous forecast updates occur is in open ended funds,
where the fund manager
-- if notified as and when share remissions and purchases are ordered during the trading day --
could pro-actively trade in the underlying assets on the trading day itself rather than on the next,
so as to increase the liquidity of the fund.
In this case the standard deviation of the updates on the volume target often follows a ``$\cup$'' profile instead.}
Further investigation is thus needed to analyse the impact of the forecasts uncertainty profile on the performance of our approach relative to the model of \citet{Almgren2001}.
Since the optimal redistribution coefficients depend on the profile of the standard deviation of the forecast updates,
these two points should be jointly addressed in further investigations.

%% file: sections/6-conclusion.tex
\section{Conclusion}%
\label{sec:conclusion}

In this paper we considered the optimal trade execution problem in a setting where price uncertainty grows in time,
but,
unlike what is assumed in the existing literature,
the required trade volume is also uncertain and becomes only known at the end of the execution period.
We assume that forecasts of increasing accuracy become available,
so that volume uncertainty decreases in time, while price uncertainty increases. 
The model presented in this paper is designed to manage both uncertainties via risk terms,
so as to be pre-computable and avoid the combinatorial explosion of dynamic programming approaches.
We have demonstrated that the model has desirable convexity properties,
guaranteeing the existence and uniqueness of a convex set of optimal strategies.
We show that it produces significantly lower transaction costs in comparison to classical trade execution approaches, 
even if the latter allow for recourse whenever a new forecast becomes available.